\documentclass[twocolumn,showpacs,preprintnumbers,amsmath,amssymb]{revtex4}

\usepackage{graphicx}
\usepackage{dcolumn}
\usepackage{bm}

\begin{document}


\title{The \mbox{\boldmath $D_{sJ}^{*}(2317)$} 
and \mbox{\boldmath $D_{sJ}(2460)$} 
mesons in \mbox{\boldmath $\tilde U(12)$}-classification scheme of hadrons}

\author{Shin Ishida}
 \altaffiliation{Senior Research Fellow}
\affiliation{%
Research Institute of Science and Technology, 
College of Science and Technology,\\
Nihon University, Tokyo 101-8308, Japan\\
}%

\author{Muneyuki Ishida}
\affiliation{%
Department of Physics, Meisei University, Hino 191-8506, Japan\\
}%

\author{Kenji Yamada and Tomohito Maeda}
\affiliation{%
Department of Engineering Science, Junior College Funabashi Campus,\\
Nihon University, Funabashi 274-8501, Japan\\
}%

\author{Masuho Oda}
\affiliation{%
Faculty of Engineering, Kokushikan University, Tokyo 154-8515, Japan\\
}%

\date{\today}

\begin{abstract}
The narrow mesons, $D_{sJ}^{*}(2317)$ and $D_{sJ}(2460)$, observed recently in the final states $D_s^+ \pi^0$ and $D_s^{*+} \pi^0$
are pointed out to be naturally assigned as the ground-state scalar and axial-vector chiral states
in the $(c\bar s)$ system, which would newly appear in the covariant $\tilde U(12)$ hadron-classification scheme
proposed a few years ago. 
We predict the comparatively large electromagnetic decay widths
to other models, which are due to the intrinsic electric dipole moment. 
The SELEX state $D_{sJ}(2632)$ is also able to be assigned to the $P$-wave chiral state 
with $J^P=1^-$ in the $\tilde U(12)$-classification scheme.  
\end{abstract}

\maketitle
\section{\label{sec:level0}
Introduction
}
\subsection{\label{sec:0A}
Present status of hadron spectroscopy and \mbox{\boldmath $D_{sJ}^{*}(2317)$} 
and \mbox{\boldmath $D_{sJ}(2460)$} 
mesons 
}
There exist the two contrasting, non-relativistic and relativistic approaches
for describing composite hadrons:
The former is based on the non-relativistic quark model(NRQM) with the approximate
$SU(6)_{SF}\times O(3)_L$ symmetry($S$, $F$ and $L$ denoting 
 Pauli-spin, flavor and orbital angular-momentum of constituent light-quarks, 
respectively) and gives a theoretical base to the particle data group(PDG) 
level-classification\cite{pdg},
while the latter is based on the field theory with the spontaneously-broken 
chiral symmetry.
It is widely accepted that $\pi$ meson octet has the property as a Nambu-Goldstone boson
in the case of spontaneous breaking of chiral symmetry.

Owing to the recent progress, both theoretical and experimental, the existence of light
$\sigma$-meson as chiral partner of $\pi$-meson 
seems to be established especially through
the analysis of various $\pi\pi$-production processes\cite{YITP}.
This gives a strong support for the relativistic approach and suggests to assign a series of 
low-mass scalar mesons as the scalar $\sigma$-nonet 
$\{ a_0(980),\sigma (600), f_0(980),\kappa (900) \}$
in the $q\bar q$ ground states.

The difficulty here of the conventional $SU(6)_{SF}
\times O(3)_L$ scheme in the framework of 
NRQM is that there are no appropriate seats for the $\sigma$ nonet in the $(q\bar q)$ ground state.
More seriously, it is, in principle, not able to treat the chiral symmetry,
and it could not refer to the well-known property of $\pi$ and $\sigma$ meson nonets as being
mutual chiral partners. In order to treat the chiral symmetry, the manifestly Lorentz-covariant
character of the relevant scheme is indispensable, since the generator
of chiral transformation for constituent quarks 
is defined as $\gamma_5\equiv\gamma_1\gamma_2\gamma_3\gamma_4$.

Recent discovery of new narrow resonant states 
$D_{sJ}^{*}(2317)/D_{sJ}(2460)$\cite{exp}
 causes further a serious problem
in hadron physics: They have the quantum numbers, $J^P=0^+/1^+$, respectively.
However, their mass values are too low to be assigned as the corresponding $(c\bar s)$
$P$-wave states in NRQM, although moderate as the $S$-wave states(, where missing
positive-parity seats). Accordingly, 
they are mostly interpreted as $4q$-states\cite{4q,CHou}, 
$DK/D^*K$\cite{Molecule} and $D_s\pi$\cite{Szc} molecular states, or 
singularities of unitarized meson-meson scattering amplitudes in 
chiral models\cite{scatt}. 

In the preceding works\cite{BEH,NRZ}, the authors presupposed the exsistence of 
{\it degenerate} two heavy-spin multiplets, 
forming a linear representation of chiral symmtry and 
consisting of $J=0$- and $1$-members, 
$H(0^{-},1^{-})$ and $H^{'}(0^{+},1^{+})$, where $H$ particles are 
assigned to the conventional ground state $P_{s}$ and $V_{\mu}$ mesons, 
and $H^{'}$ particles are to the relevant scalar $S$ and axial-vector $A_{\mu}$ mesons, 
respectively. 
Then the authors\cite{BEH} derived the interesting results on the mass-splitting 
between the chiral partners $H$ and $H^\prime$.
However, the identification of these particles $H^{'}(0^{+},1^{+})$ in the 
level-classification scheme is obscure, although the authors, in application to 
radiative decay processes, assigned the $H^{'}$ to the P-wave $j_{q}^{P}={\frac{1}{2}}^{+}$ 
multiplet in the heavy quark effective theory(HQET).

A few years ago we have proposed a manifestly covariant scheme, 
the $\tilde U(12)$-classification scheme\cite{rf2}, 
which maintain on the one hand the successful part of 
$SU(6)_{SF}\times O(3)_{L}$ scheme and, on the other hand, it reconciles
the quark model to the chiral symmetry. 
It has a unitary symmetry in the hadron rest frame, ``static $U(12)$'', 
embedded in the 
covariant $\tilde U(12)$-tensor space(see, the next-subsection). 

The static $U(12)$ includes both its subgroup, 
$SU(6)_{SF}$ and chiral $SU(3)_L \times SU(3)_R$, as $U(12) \supset 
SU(6)_{SF}
\times SU(2)_{\rho}$ on the one hand, and as $U(12) \supset SU(3)_{L} \times SU(3)_{R} 
\times SU(2)_{\sigma}$ on the other hand. 
The $SU(2)_{\rho}$ and $SU(2)_{\sigma}$ are Pauli's spin groups 
concerning the boosting and the intrinsic-spin rotation, respectively, 
of the constituent quarks, being connected with decomposition of Dirac 
$\gamma$-matrices, $\gamma \equiv \rho \otimes \sigma$.
The freedom on $SU(2)_{\rho}$, which is indispensable for covariant description of 
spin one-half particles, plays also an important role to define 
the rule of chiral transformation for general quark-composite hadron. 
This becomes possible by introduction of chiral spinors into the complete set 
of fundamental Dirac-spinors of the $\tilde{U}(12)$ tensor space, as will be 
explained in the following sub-section.

The purpose of this work is to investigate the properties of the controversial
$D_{sJ}^*(2317)$ and $D_{sJ}(2460)$ mesons in the framework 
of our new scheme. 
The heavy-light-quark meson system is 
the most suitable for testing the validity of our scheme, 
since it is expected to show clearly both the non-relativistic and relativistic behaviors
concerning the heavy and light constituent quarks, 
that is, the heavy quark symmetry(HQS) and the light quark chiral symmetry, 
respectively.

In our scheme, the degenerate, two heavy-spin multiplets 
$H(0^{-},1^{-})$ and $H^{'}(0^{+},1^{+})$ above mentioned 
are assigned to be the ground states $(c\bar{s})$ mesons, 
;where $H$ and $H^{'}$, 
being the eigen-states of $\rho_{3}$-spin(concerning the light constituent quarks) 
with the eigen-values, $r=+$ and $-$, respectively, 
become mutually chiral partners, while their members belong to the multiplets 
with $S=(0,1)$ of $SU(2)_{\sigma}$.

Thus, in our covariant classification scheme, the relevant 
{\it $D_{sJ}^{*}(2317)/D_{sJ}(2460)$ are naturally assigned as 
the scalar$/$axial-vector chiralons, 
playing the role of chiral partners of pseudo-scalar$/$vector Paulons, 
$D_s(1968)/D_s^*(2112)$, in a linear representation of chiral symmetry 
for light quarks.
All these particles are assigned as the S-wave $(c\bar s)$ ground states,} 
and the ordinary P-wave $j_q^P=\frac{1}{2}^+$ multiplet with 
$J^{P}=(0^{+},1^{+})$ is expected to exist 
as different particles in the slightly-higher mass region. (see, the discussion below 
Eq.(\ref{EEq24}).)

Furthermore, it will be shown that their properties predicted from our scheme
are almost consistent with the present experiments.
\subsection{\label{sec:0B}
Lorentz-covariance and overlooked freedom \mbox{\boldmath $SU(2)_{\rho}$} of 
composite hadrons}
In this sub-section we recaptulate fundamental ideas and basic-points 
of our scheme, which is described rigorously in the sections \ref{sec:level1} 
and \ref{sec:Cov descr}.
In our scheme we start from the master Klein-Gordon(K.G.) equation on 
the wave function(WF) of composite hadrons with the 
{\it squared-mass} ${\cal M}^2 (r_{\mu})$ operator, acting on the 
Lorentz-space
$O(3,1)_{L}$ for relative space-time coordinates $r_{\mu}$'s of constituents.
Here we should like to mention that the observable entity is not the 
{\it confined quarks but the comosite hadrons}.

Aside from the center of mass(CM) plane-wave motion, the internal WF of hadrons 
are given, in the ideal limit, as eigen-functions of the 
${\cal M}^2 (r_{\mu})$, which is taken to be the covariant oscillator of 
Yukawa-type\cite{rfY}. 
Their spectra simulates, by imposing a subsidiary condition\cite{Takabayashi} to 
freeze the relative-time freedom, those of oscillator on $O(3)_{L}$ in NRQM, 
and the WF represents the Lorentz-contraction effects\cite{FujimuraKobayashiNamiki}
 due to CM motion.

Further, the WF of hadrons are set up(for a while considering light-quarks) 
to be appropriate tensors in $\tilde{U}(12) \times O(3,1)_{L}$-space, 
reflecting the quark-comosite structure of relevant hadrons. 
Here $\tilde{U}(12)$ includes the subgroups $\tilde{U}(12) \supset 
\tilde{U}(4)_{DS} \times SU(3)_{F}$($\tilde{U}(4)_{DS}$ being the 
pseudo-uitary homogeneous Lorentz-group ${\cal L}_{4}$ for the Dirac spinors).

Then the spin-flavor WF of relevant hadrons become covariant tensors 
in the $\tilde{U}(12)$-space with definite four-velocity 
$v_{\mu}\equiv P_{\mu}/M$($P_{\mu}$($M$) 
being the four momentum(mass) of the hadrons) 
%

The spin WF of hadrons are tensors 
in $\tilde{U}(4)_{DS}$ space, and represented by relevant multi-product of 
the original Dirac spinors (as fundamental vectors in 
$\tilde{U}(4)_{DS}$-tensor space), $W_{\alpha}{}^{(\pm)}(v)$ and its 
Pauli-conjugates $\overline{W^{\beta}{}^{(\pm)}}(v)$. 
Here, $W_{\alpha}{}^{(+)}(v)$ and $W_{\alpha}{}^{(-)}(v)$ denote 
positive and negative frequency parts, respectively, of the local Klein-Gordon 
equations for spin one-half particles corresponding to constituent-quarks.:
\begin{eqnarray}
W_\alpha (X) &\equiv& \sum_{P_{\mu}(P_{0}>0)}
(e^{iP\cdot X} W_{\alpha}^{(+)} (P)
+e^{-iP\cdot X} W_{\alpha}^{(-)} (P)),\nonumber \\
W_{\alpha}^{(+)} (P)&=&\{ u_{r=+,s,\alpha}(P),\ v_{\bar{r}=-,\bar{s},\alpha}(P)
;s,\bar s=(+,-) \},\nonumber\\
W_{\alpha}^{(-)} (P)&=&\{ v_{\bar{r}=+,\bar{s},\alpha}(P),\ u_{r=-,s,\alpha}(P),\ 
;\bar{s},s=(+,-) \}.\nonumber\\
-(\gamma_\mu \partial_\mu &+& M ) W(X)=0\ .
\label{eqInt}
\end{eqnarray}
Both $W_{\alpha}^{(+)}(P)$ and $W_{\alpha}^{(-)}(P)$ consist, respectively, of 
four members as above, 
where $r(\bar{r})=\pm$ and $s(\bar{s})=\pm$ denote the eigen-values of 
$\rho_{3}(\bar{\rho}_{3})$ and $\sigma_{3}(\bar{\sigma}_{3})$ in the rest 
frame of hadrons($\mbox{\boldmath $v$}=\mbox{\boldmath $0$}$)(see 
Eqs.(\ref{eq:localKG}) through (\ref{eq:explicit-form})).

The positive- and negative-frequency parts of the hadron WF $\Phi(X,r)$(its 
Pauri-conjugate $\bar{\Phi}$) are second-quantized as annihilation(creation) 
operator and creation(annihilation) operators of relevant hadrons(its 
anti-particles), in conformity with the conventional crossing rule for hadrons.

The meaning of the static $U(12)$, 
embedded in the 
covariant $\tilde U(12)$-tensor space(here, $U(4)_{DS}$ 
embedded in the $\tilde{U}(4)_{DS}$)
, mentioned in the last 
sub-section, is that the $\tilde{U}(4)_{DS}$-invariant 
squared-mass term in action integral become 
static $U(4)$-symmetric by inserting a Lorentz-invariant factor, 
the unitarizer $F_{U}(X)$, 
between the trace of 
light-quark spin-indices(see Eq.(\ref{action}))).

Here it is to be remarked on the important fact which has been thus far 
overlooked: conventionally the spinors $u_{-,s}$($v_{-,\bar{s}}$) for 
quarks(anti-quarks) are identified with the spinor $v_{+,\bar{s}}$($u_{+,s}$) 
for anti-quarks(quarks), being based on the hole-theory in the case of free 
quarks. However, it is not applicable to the confined quarks, coexisting with 
the other quarks(see \ref{sec:1B}), and all the above four-members, separately, of 
$W_{\alpha}^{(+)}(v)$ for quarks and of $W_{\alpha}^{(-)}(v)$ for anti-quarks 
are required as the members of complete set of fundamental vectors to 
describe the spin WF of hadrons. This is one of the corner-stones of our 
scheme, and comes from our application of Klein-Gordon equation as the master 
equation, describing the WF of {\it observable hadrons}.

This fact is shown as follows: we start from the K.G. equation 
for Dirac spinor $\psi (x)$ with four-components as
\begin{eqnarray}
(\Box - \kappa^{2})\psi (x)
=(\gamma_{\mu}\partial_{\mu}+\kappa)(\gamma_{\mu}\partial_{\mu}-\kappa)
 \psi (x)=0.
\end{eqnarray}
This has the two types of solution, which satisfy, respectively, the following 
Dirac-type equations as
\begin{eqnarray}
(\gamma_{\mu}\partial_{\mu}+\kappa)\psi_{+}(x)=0, 
(\gamma_{\mu}\partial_{\mu}-\kappa)\psi_{-}(x)=0.
\end{eqnarray}
As is clearly seen from these equations, the $\psi_{+}$ and $\psi_{-}$ form 
a chiral doublet, whose spinor wave functions, $\psi_{+,\alpha}(v)$ and 
$\psi_{-,\alpha}(v)$, consist of totally the eight ($8=4 \times 2$) 
independent-ones, and are evidently 
equivalent to \{ $u_{r=+,s,\alpha}(v),
v_{\bar{r}=+,\bar{s},\alpha}(v)$\} and \{ $u_{r=-,s,\alpha}(v),
v_{\bar{r}=-,\bar{s},\alpha}(v)$\}, respectively. 

Here, it may be worthwhile to note that the above reasonning led to the chiral doublet, 
$\psi_{+}$ and $\psi_{-}$, is similar to that in the case of deriving the 
chiral-invariant weak-currents of V-A type\cite{Feyn-Galletc}.

The freedom on $SU(2)_{\rho}$, concerning discrimination between the 
eigen-states of $\rho_{3}$-spin of fundamental vectors $W_{\alpha}^{(\pm)}$'s 
of $\tilde{U}(4)_{DS}$-tensor space(which might be vanished in the case 
of not-confined constituent-quarks) plays important roles in composite hadrons 
as follows:

(i) In defining the booster for special Lorentz-transformation, 
the $SU(2)_{\rho}$ space is indispensable as
\begin{eqnarray}
S_{B}(\mbox{\boldmath $P$})&=&e^{-i\mbox{\boldmath $b$} \cdot \mbox{\boldmath $K$}}, 
K_{i}=\frac{i}{2}\sigma_{i} \otimes \rho_{1}.\nonumber \\
(\mbox{\boldmath $b$}& \equiv & \hat{\mbox{\boldmath $v$}} \cosh^{-1}v_{0}) .
\end{eqnarray}

(ii) The transformation for reflection of 
space-time coordinates is given, as
\begin{eqnarray}
&&X_{\mu} \to X_{\mu}^{'}=-X_{\mu}; ~~~~~~~~~~~~~~~~~~~~~~~~~\nonumber \\
&&W_{\alpha}(X) \to W_{\alpha}^{'}(X^{'})=W_{\alpha}^{'}(-X)=
(-\gamma_{5}W(-X))_{\alpha},\nonumber \\
&&-\gamma_{5} W_{{\pm},s}(v_{\mu})=\rho_{1}W_{\pm,s}(v_{\mu})=
W_{\mp,s}(v_{\mu}).
\end{eqnarray}

(iii) The transformation for reflection of space coordinates are given, as
\begin{eqnarray}
&&X_{i} \to  X_{i}^{'}=-X_{i};X_{0}^{'}=X_{0};\nonumber \\
&&W_{\alpha}(X) \to W_{\alpha}^{'}(-X_{i},X_{0})=
(\gamma_{4}W(-X_{i},X_{0}))_{\alpha},\nonumber \\
&&\gamma_{4} W_{{\pm},s}(v_{i}=0)=\rho_{3}W_{\pm,s}(v_{i}=0)=
\pm W_{\pm,s}(v_{i}=0).\ \ \ \ \ \ \ \  
\end{eqnarray}

The above fact (iii) means that the 
$W_{+,s}$ and $W_{-,s}$ 
form a parity-doublet, 
while the fact (ii) implies that (generator of) 
chiral transformation $-\gamma_{5}$ transforms the members of this doublet each other. 

Above we have given only the rule for relevant transformations of fundamental 
vectors $W_{\alpha}^{(\pm)}$'s in the $\tilde{U}(4)_{DS}$-tensor space.
The rules of any transformation on the spin-flavor WF of composite hadrons, 
tensors in $\tilde{U}(12)$-space, are derived from those of constituents as 
fundamental vectors of this space. 

In this connection we note especially, 
concerning our relevant H/L meson system, that the static $U(12)$ symmetry 
expected on the constituent L-quarks, leads to existence of parity-doublets
(due to (iii)), whose members play mutually the role of chiral partners(due 
to (ii)).

Finally here, on the basis of above arguments, we give some comments on the 
conventional approach; applying Dirac equation to light-quarks in the 
heavy-light(HL) quark meson\cite{RN1,RN2}, and on the covariant kinematical 
framework\cite{RN3} based on heavy-quark(HQ) 
effective theory:These preceding works are surely appropriate for 
introducing HQ symmetry in the framework. 
However, the framework of Ref.\cite{RN1,RN2} is not covariant, since 
the whole hadron seems to be at rest($\mbox{\boldmath $v$}=
\mbox{\boldmath $0$}$), and the booster becomes, due to (i), identity.
In the work\cite{RN3} the framework is covariant but missing there 
the definition of chiral symmetry. 
The framework of Ref.{\cite{RN3}} is equivalent to that of our preceding scheme, 
the oscillator quark model\cite{Sakura,urciton}, before introducing the 
``chiral spinors'', $u_{-,s}$ and $v_{-,\bar{s}}$, into the members of 
complete set of fundamental vectors $W_{\alpha}^{(+)}(P)$ and 
$W_{\alpha}^{(-)}(P)$.


\section{\label{sec:level1}
Essentials and origin of \mbox{\boldmath $\tilde U(12)$}-classification scheme
}
Before going into detailed application of our scheme which might be unfamiliar to most 
readers, we shall describe the essential points and the 
origin of the $\tilde U(12)$ scheme, which has a long history.
(As for details, see our review articles\cite{rf1} and the original ones\cite{rf2}.)

\subsection{\label{sec:1A}
Covariant \mbox{\boldmath $\tilde U(12)$}-classification scheme and static \mbox{\boldmath $U(12)$} symmetry
}
The framework of $\tilde U(12)$-classification scheme is manifestly Lorentz-covariant, 
and the hadron wave functions (WF)
are supposed to be generally tensors 
in the $\tilde U(12)_{SF}\times O(3,1)_L$ space, where the 
$\tilde U(12)_{SF}\supset U(3)_F \times \tilde U(4)_{DS}$. 

Here it is to be noted that we do not assume the $\tilde U(12)$ symmetry or
any other rigorous relativistic symmetry
(including the generators of Lorentz transformation $\Sigma_{\mu\nu}$ as symmetry generators), 
leading to such an irreducible
representation as containing its members with different spins.
There is no such type of relativistic symmetries because of No-Go theorem by
Coleman and Mandula\cite{ColemanM}.
Instead we mean by the term of $\tilde U(12)$-classification scheme that
all members belonging to the same multiplet have the same
{\it{squared-mass in the ideal limit}}
(, where are neglected the effects of perturbative QCD and those due to the 
spontaneously-broken chiral symmetry)
 and that strong interactions
are consistent with a unitary group of the $U(12)_{SF}$ symmetry, when all relevant hadrons
are at rest.
In this sense our scheme is, more strictly, to be called the static $U(12)$ symmetry 
scheme.
Because of this rest condition\cite{Roman}, 
No-Go theorem is not applicable to our classification scheme,
since static $U(12)$ does not include $\Sigma_{\mu\nu}$ 
as its generators\cite{foot1}. (As for details, see the appendix.)

In our scheme the representation space for light constituent quarks 
is extended from the non-relativistic (NR) $SU(6)_{SF}\times O(3)_L$ space to
the covariant $\tilde U(12)_{SF}\times O(3,1)_L$ space: 
This implies the introduction of new additional $SU(2)$ space on the 
$\rho$-spin and mean the extension of conventional
 $SU(6)_{SF}$ to $U(12) \supset 
SU(6)_{SF} \times SU(2)_{\rho}$. 
In another representation, the static $U(12)$ also includes the conventional, 
chiral $SU(3)_{L}\times SU(3)_{R} \times SU(2)_{\sigma}$ as a subgroup. 
Here the $\rho$-spin and the Pauli's $\sigma$-spin are those 
concerning with the decomposition of Dirac $\gamma$ matrices, 
$\gamma \equiv \sigma {\scriptstyle \bigotimes} \rho$.
In the Pauli-Dirac representation the direction of $\rho_3$-spin denotes that of time-flow
in the rest frame of relevant hadron.
These situations are shown in table {\ref{tabSym}}.
\begin{table}
\caption{Static $U(12)$ symmetry and its covariant representation-space
 for light quarks}
\begin{center}
\begin{ruledtabular}
\begin{tabular}{ll}
  \underline{\rm Symmetry} & \underline{\rm Representation\ Space}\\
hadron at rest($\mbox{\boldmath $P$}=0$)&hadron in moving frame
($\mbox{\boldmath $P$}$)\\
$U(12)_{SF} \times O(3)_{L}$&$\tilde{U}(12)_{SF}\times O(3,1)_{L}$\\
$U(12)_{SF} \supset  
      SU(6)_{SF} \times SU(2)_\rho $,\\
$\supset 
      SU(3)_{L} \times SU(3)_{R} \times SU(2)_\sigma $
\end{tabular}
\end{ruledtabular}
\end{center}
\label{tabSym}
\end{table}

\subsection{\label{sec:1add}
Origin of $\tilde U(12)$-classification scheme
}
The origin of $\tilde U(12)$-classification scheme is traced back long ago.
In 1965, shortly after the proposal of quark model and, successively, 
of the $SU(6)_{SF}$ symmetry,
Salam et al. and Sakita-Wali independently 
proposed\cite{Delbourgo} the $\tilde U(12)$ symmetry
as a relativistic extension of the $SU(6)_{SF}$ symmetry.
However, it is now well-known that its relativistic extension 
as a rigorous mathematical group
is impossible, as was mentioned in the preceding subsection.

In 1970 one of the authors proposed the urciton scheme\cite{urciton}, 
which is the direct origin of our new scheme, for the purpose of treating multi-quark
hadrons systematically and covariantly.
Here the constituent quarks are regarded as excitons related to the relevant hadron,
and the hadron WF correspond to the Fock amplitudes for the system of
multi-exciton quarks, moving with the same velocity as the relevant hadron.
As a result we can show that the hadron WF reproduces the successful contents
(that is, the results of $SU(6)_{SF}$) of the original $\tilde U(12)$ symmetry\cite{Delbourgo}. 
(See, Appendix 3.)
The WF in  this scheme is manifestly covariant.
However, we can not yet treat the chiral symmetry, since it was assumed there that
only ``boosted-Pauli spinors'' are applied as physical ones 
(following the original $\tilde U(12)$-papers\cite{Delbourgo}). 
Our new $\tilde U(12)$-classification scheme is developed only by discarding
this now-unnecessary restriction and by taking into account all elements
of the complete set to be physical states in expanding $\tilde U(12)$ (tensor)-space.

\subsection{\label{sec:1B}
Fundamental representation of confined-quarks and 
chiral states$/$chiralons
}
Because of the new freedom $SU(2)_{\rho}$, 
our scheme becomes reconcilable with the chiral symmetry, 
and we are led to existence of new states for hadrons
(named chiral states), 
which are out of the conventional NR scheme.
The fundamental vectors in $\tilde{U}(4)_{DS}$ tensor-space are given by
the original Dirac spinors, $W_{\alpha}{}^{(\pm)}(P)$ and its Pauli-conjugate
 $\overline{W^{(\pm)}(P)}^{\beta}\equiv (W^{(\pm)}(P)^{\dagger}\gamma_{4})
^{\beta}$. 
Here $W_{\alpha}{}^{(+)}(P)$ and $W_{\alpha}{}^{(-)}(P)$
 are the four-dimentional 
Fourier amplitudes of $W^{(+)}(x)$ and $W^{(-)}(x)$, respectively, 
which are the 
positive and negative frequency parts of the solutions of the 
``local'' Klein-Gordon equation $W_{\alpha}(x)$, $x_{\mu}$ being 
the space-time coordinate in $O(3,1)_{L}$ (see section \ref{sec:C}).
The complete set of $W^{(+)}(P)$ ($W^{(-)}(P)$) consists of the four elements
$u_{r,s}(P)$($v_{\bar{r},\bar{s}}(P)$)
 describing the spinor freedom of 
confined quarks(anti-quarks) inside of hadrons, where 
$r(\bar{r})=\pm$ and $s(\bar{s})=\pm$ 
denote eigen values of $\rho_{3}$($\bar{\rho}_{3}=-\rho_{3}^{t}$) 
and $\sigma_{3}$($\bar{\sigma}_{3}=-\sigma_{3}^{t}$) for $u$($v$), respectively, 
and $P_{\mu}$($P_{0} \equiv \sqrt{\mbox{\boldmath $P$}^2+M^2}>0$) being the 
center of mass four momenta of hadrons (not quarks): 
Conventionally, 
 the spinors $u_{-,s}$($v_{-,\bar{s}}$) for quarks(anti-quarks)
 are identified with the spinors $v_{+,\bar{s}}$($u_{+,s}$)
  for anti-quarks(quarks).
 This is based upon the hole-theory on the free quark field theory.
 Correspondingly, in NRQM only the NR two-component Pauli-spinors 
 $\chi_{s}$($\chi_{\bar{s}}$) for quarks(anti-quarks), 
 which becomes equivalent to
 the upper(lower) two-components of four-component boosted-Pauli spinors
 $u_{+,s}$($v_{+,\bar{s}}$) in the static limit with
 $\mbox{\boldmath $P$}=\mbox{\boldmath $0$}$, are applied.
 In the original $\tilde{U}(12)$-symmetry scheme (, and in the previous
 framework of covariant oscillator quark model(COQM)\cite{Sakura},)
  only the boosted-Pauli spinors
 $u_{+,s}$($v_{+,\bar{s}}$) are regarded as physical states. 
 
 However, 
 the above picture on hole theory and the identification of $u_{-,s}
 =v_{+,\bar{s}}$
 ($v_{-,\bar{s}}=u_{+,s}$) is only applicable to the free quarks 
 (or to whole free-hadrons), and not separately 
 to the confined constituent-quarks,
 coexisting with the other quarks.
 This is so, because the application of hole-theory implies that all quantum numbers 
 of particle-hole are to be replaced by their conjugates. 
 Especially, concerning the color freedom, the application induces the change of 
 \mbox{\boldmath $3_{c}$} of quark-holes to 
 the \mbox{\boldmath $3^{*}_{c}$} of anti-quarks, leading to the violation of 
 color-singlet condition for the relevant hadron. 
  Accordingly, in describing the spinor
 WF of composite hadrons covariantly, all four Dirac spinors, 
 $u_{\pm,\pm}$ and $v_{\pm,\pm}$, respectively, for quarks and
 for anti-quarks, are required as the elements of complete set
 of the fundamental vectors in $\tilde{U}(4)_{DS}$-space.
 
 As is described above, our Dirac spinors as the fundamental vectors in
  $\tilde{U}(4)_{DS}$ (tensor)-space is representing some mathematical 
  quantity, to be appropriately called ``exciton quarks'', 
  simulating the properties of constituent quarks inside hadrons, 
  and different from those representing constituent-quarks in the dynamical 
  composite models. We shall call our Dirac spinors as the ur-citon spinor. 
  The notion of exciton quarks (and its name urciton) 
  was first introduced in the paper\cite{urciton}
  long ago, and the urciton spinor seems to be of the similar nature to 
  the u-spinor, 
  introduced in ``the 144-fold way out'' from the trouble of relativistic 
  $SU(6)$ \cite{Weinberg}. (See, Appendix A.6.)
 
 These urciton spinors are transformed with each other by operating 
 $-\gamma_{5}$, the generator of chiral transformation, as 
 $u_{+,s} \leftrightarrow u_{-,s}$, $v_{+,\bar{s}}
  \leftrightarrow v_{-,\bar{s}}$
 , and play mutually a role of chiral partners. For later convenience 
 we call the spinors $u_{+,s}$ and $v_{+,\bar{s}}$
  the (relativistic) Pauli-spinors ;
 while do $u_{-,s}$ and $v_{-,\bar{s}}$ the chiral-spinors 
 (see, \ref{sec:Cov descr} as for
 the strict formula).
 
The above mentioned existence of new states for hadrons,
 to be called chiral states,
 is due to
 this introduction of $u_{-,s}$ and $v_{-,\bar{s}}$ in representing the
 confined quarks inside of hadrons: 
The chiral states of hadrons are defined as being represented by 
the tensors containing at least one 
 $\tilde{U}(4)_{DS}$-index of
 chiral-spinors ,while we define, the Pauli states of hadrons 
 as being described by those of only Pauli-spinors. 
 We call, especially, the hadrons being represented purely by the 
 Pauli/chiral states as Paulons/chiralons. 
 Here it should be noted that the physical hadrons generally belongs to a 
  superposition of the Pauli- and chiral-states.
They, Paulons and chiralons, form 
a linear representation of chiral symmetry.

Furthermore, it should be remarked that due to the introduction\cite{rf1}
of chiral states into physical complete set of ${\cal S}$-matrix bases,
our $\tilde U(12)_{\rm stat}$-symmetry classification scheme becomes
free from the problem of unitarity\cite{Weinberg} in the original $\tilde U(12)_{SF}$-symmetry
scheme\cite{Delbourgo}. (See the Appendix A.6 for details.)

\subsection{\label{sec:1B-newD}
Level-structures of light-quark mesons and baryons, and heavy-light quark mesons}

The systematic and rather rigorous considerations of the general meson-system 
in our new scheme have been given in the second paper of Ref.\cite{rf2},
and the level structure of light-quark meson and baryon system have been shortly described
in the first paper of Ref.\cite{rf2}. 
Here we will recapitulate the essential points and pick up the candidates for chiralons.

The light-light(LL) quark mesons in the ground-state($L=0$),
which are classified as ${\bf 6}\times {\bf 6}^* = {\bf 36}$ in $SU(6)_{SF}$,
are assigned as  
${\bf 12}\times {\bf 12}^* = {\bf 144}$
in $\tilde U(12)_{SF}$.
The {\bf 144} includes two sets
of pseudoscalar and vector nonets, $\{ P_s^{(N)},V^{(N)}; P_s^{(E)},V^{(E)} \}$
and also two sets of scalar and axial vector nonets, $\{ S^{(N)},A^{(N)};S^{(E)},A^{(E)} \}$.
The respective members in the two brackets play a role of chiral partners
mutually.
The $\pi$-meson nonet and $\sigma$-meson nonet are assigned to the $P_s^{(N)}$ and $S^{(N)}$,
respectively, where the former being maximally mixed states of the 
Pauli- and chiral-states, while the latter being purely-chiral 
states(that is, chiralons).
The conventional $\rho$-meson nonet is conjectured to be dominantly the Pauli 
states(that is, Paulons).
Concerning the remaining nonets, $P_s^{(E)}$, $V_\mu^{(E)}$ and $S^{(E)}$, $A^{(N)}$, $A^{(E)}$,
it is suggestive that the identification of the relevant $P$-wave states in the conventional scheme
is still in some confusion, and that the lower excited-vector and pseudo-scalar states
seem to contain some extra-levels.

The light quark $(qqq)$ baryon system in the ground $S$-wave states is classified as
$({\bf 12}\times{\bf 12}\times{\bf 12})_S={\bf 364}$, which includes
baryon and anti-baryon. The {\bf 182} of baryons is decomposed into
${\bf 182}={\bf 56}+{\bf 70}+{\bf 56}^\prime$, where
{\bf 56} corresponds to the conventional {\bf 56} in $SU(6)_{SF}$.
Additional {\bf 70}$({\bf 56}^\prime )$ with negative(positive) parity 
have generally very wide widths 
and are considered to be observed only as backgrounds, except for the cases of
the problematic $\Lambda (1405)$ (Roper $N(1440)$).
(As for more details, see Appendix A.3.)

Inclusion of heavy quarks ($Q$) in  our covariant $\tilde{U}(12)$-classification
 scheme is straightforward: In table {\ref{tabSym}} the representation space 
 $\tilde{U}(12)_{SF}$ for the flavor and spinor freedom of light-quarks
 is to be extended to the $[\tilde{U}(12)_{SF}]_{q} \otimes 
[\tilde{U}(4)_{SF}]_{Q}$ for the heavy-light(HL) quark hadrons, the 
 $[\tilde{U}(4)_{SF}]_{Q}$ being $[U(1)_{F}\otimes\tilde{U}(4)_{DS}]_{Q}$.
  The symmetry in the rest frame becomes  $[U(12)_{SF}]_{q}\otimes
[U(2)_{SF}]_{Q}$, where the $U(2)_{SF}$ being $U(1)_{F}\otimes SU(2)_{S}$,
 since for the heavy quarks inside of hadrons the heavy quark spin-symmetry 
is valid, implying only the boosted Pauli-spinors are required 
as ur-citon spinors.

The HL quark mesons in the ground state (L=0), which are classified as 
$\mbox{\boldmath $6$}\times \mbox{\boldmath $2$} = \mbox{\boldmath $12$}$ 
in the non-relativistic $SU(6)_{q}\times SU(2)_{Q}$ scheme, are assigned as 
$\mbox{\boldmath $12$}\times \mbox{\boldmath $2$} = \mbox{\boldmath $24$}$ 
in te static $U(12)_{q}\otimes SU(2)_{Q}$ symmetry scheme (, embeded in the 
covariant $[\tilde{U}(12)_{SF}]_{q}\times
[\tilde{U}(4)_{SF}]_{Q}$ representation space ). The $\mbox{\boldmath $24$}$ 
includes the $\mbox{\boldmath $12$}$ of 
the conventional pseudo-scalar $0^{-}$- and vector 
$1^{-}$ $SU(3)_{F}$ triplets 
and the $\mbox{\boldmath $12$}$ of the newly-appearing scalar $0^{+}$-
 and axial-vector $1^{+}$-$SU(3)_{F}$ triplets in the ($q\bar{Q}$) system. 
The formers are Paulons, 
while the latters are chiralons. 
Their anti-particles in the ($\bar{q}Q$) system belong to the multiplet 
$\mbox{\boldmath $12^{*}$} \times \mbox{\boldmath $2$}=\mbox{\boldmath $24^{*}$}$.
\section{Covariant description of heavy-light(HL) quark mesons}
\label{sec:Cov descr} 
In this section we shall review briefly on how to describe covariantly
 the composite hadrons, in so far as concerned with the HL-quark mesons
in the $\tilde U(12)$-classification scheme.
Firstly it should be noted that the $\tilde U(12)$-classification scheme is, in the present stage,
a mere kinematical framework proposed semi-phenomenologically for describing covariantly
composite hadrons, although it is expected to be derived dynamically from non-perturbative
treatment of QCD.
\subsection{\label{sec:2A}
Attributes and wave functions of composite hadrons
}
Our relevant HL mesons(, more generally composite hadrons,) should have, as their indispensable
attributes, i) definite mass and ii) spin, iii) definite Lorentz-transformation properties,
and iv) definite quark-composite structures. 
Therefore, their wave function(WF) should represent them evidently, 
and have the symmetry expected for the relevant mesons as a composite system 
(the attribute iv)), being bound by QCD.

Accordingly, we set up the wave function(WF)
of the relevant HL ($Q\bar q$)-meson (and LL ($q\bar q$)-meson) system as
\begin{eqnarray}
\Phi_A{}^B (x,y); &\ \ &  A=(\alpha ,a),\  B=(\beta ,b); \label{eq6}
\end{eqnarray}
where
$\alpha$, $\beta = (1\sim 4); a,b=(u,d,s)$ and $ (c,b)$ for $q$ and $Q$; 
 $\alpha$($\beta$) denotes the suffix of Dirac spinor of quark(anti-quark).
(The color and flavor indices, which are trivial for the relevant problem,
are omitted throughout this paper except for necessary places.) 
In setting this we have imaged as a guide the field theoretical expression for the WF as
\begin{eqnarray}
\Phi_{M,A}{}^{B}(x,y) &\sim& \langle 0 | \psi_A(x)\bar\psi^B(y) | M \rangle 
                  \nonumber\\
                  &&~~~~~+ \langle M^c | \psi_A(x)\bar\psi^B(y) | 0 \rangle , 
\label{eq:eqField}
\end{eqnarray}
where $\psi_A (\bar\psi^B)$ denotes the quark field (its Pauli-conjugate)
and $| M \rangle (| M^c \rangle )$ denotes 
the composite meson (its charge-conjugate) state.
We have also imaged, for WF of its charge-conjugate meson system, the 
field theoretical expression as 
\begin{eqnarray}
\Phi_{M^{c},B}{}^{A}(y,x) 
&\sim& \langle 0 | \psi_B(y)\bar\psi^A(x) | M^{c} \rangle 
                  \nonumber\\
                  &&~~~~~+ \langle M | \psi_B(y)\bar\psi^A(x) | 0 \rangle , 
\label{eq:eqconjField}
\end{eqnarray}
Then, the WF $\Phi$ and their Pauli-conjugates $\bar{\Phi}$, 
(defined by $\bar{\Phi}\equiv \gamma_{4}\Phi^{\dagger}
\gamma_{4}$, ) satisfy mutually the relations 
\begin{eqnarray}
\Phi_{M^{c},B}{}^{A}(y,x)=\overline{\Phi_{M}}_{B}{}^{A}(x,y),\nonumber\\
\Phi_{M,A}{}^{B}(x,y)=\overline{\Phi_{M^{c}}}_{A}{}^{B}(y,x),\label{eq:cc}
\end{eqnarray}
These relations imply that the total WF 
~$\Phi_{A}{}^{B}(x,y)$ of the composite 
meson and its charge conjugate meson system satisfy(, as they should,) the 
self-conjugate relation : 
\begin{eqnarray}
\Phi_{A}{}^{B}(x,y)=\overline{\Phi}_{A}{}^{B}(y,x),
\end{eqnarray}
where
\begin{eqnarray}
\Phi_{A}{}^{B}(x,y) \equiv \sum_{M} \Phi_{M,A}{}^{B}(x,y)
=\sum_{M^{c}} \Phi_{M^{c},B}{}^{A}(y,x).
\end{eqnarray}
\subsection{\label{sec:2B}
Klein-Gordon equation
}
In order to fix the mass of HL-mesons (the first attribute) our WF $\Phi_A{}^B$
are assumed to satisfy the master Klein-Gordon equation of Yukawa-type\cite{rfY}.
\begin{eqnarray}
\left[  \left( \partial / \partial X_\mu \right)^2 
- {\cal M}^2(r_\mu ) \right] 
 \Phi_A{}^B (X,r)   &=& 0\ ,   
\label{eqYukawa}
\end{eqnarray}
where $X_\mu$ is the CM coordinate of meson and 
$r_\mu$ is the relative coordinate.
Here the squared-mass operator ${\cal M}^2$ 
is Lorentz-scalar and assumed to be diagonal
on and independent of flavor-spinor indices, 
$A$ and $B$, of light-quarks in the ideal limit(, neglecting
possible effects due to perturbative QCD and vacuum condensate).
This assumption leads to the $U(12)_{stat}$-symmetric 
{\it squared-mass} spectra of hadrons
in the ideal limit and makes {\it our scheme reconcilable
 with the chiral symmetry} concerning the 
light quarks, as is explained in the sub-section \ref{sec:F}. 
As a concrete model of ${\cal M}^2$ we adopt the covariant oscillator
 in COQM\cite{foot3}.
\subsection{\label{sec:C}
Internal WF with definite mass and spin
}
The total WF are separated into the two(positive or negative frequency) parts 
concerning the CM plane-wave motion (with four-momentum $P_{N,\mu}$), and expanded in
terms of mass eigenstates concerning the internal space-time variables, as
\begin{widetext}
\begin{eqnarray}
{\Phi}_{A}{}^{B} (X,r) &=& \sum_{N,P_N(P_{N,0}>0)}[ 
  e^{iP_N \cdot X}\Psi_{N,A}^{(+) B}(P_N,r) + e^{-iP_N \cdot X}
   \Psi_{N,A}^{(-) B}(X,r) ] ,\nonumber\\
 && {\cal M}^2 \Psi_{N,A}^{(\pm) B}(P_N,r) = M_N^2 \Psi_{N,A}^{(\pm) B}
 (P_N,r) .
\label{eq23}
\end{eqnarray}
\end{widetext}
The internal WF $\Psi_{N,A}^{(\pm) B}$ of hadrons with definite mass $M_N$ and total spin $J=L+S$,
being tensors in the $\tilde U(4)_{D.S.}\times O(3,1)_L$ space, is given by a
relevant linear-combination of direct products of respective subspace eigen-functions, the extended 
Bargmann-Wigner(BW) spinors $W_{A}{}^{B}(P_{N})$ 
on the
$\tilde U(12)$ space and the Yukawa oscillator function $O(P_N,r)$
on the $O(3,1)_{L}$ space, as
\begin{eqnarray}
\Psi_{J,A}^{(\pm) B} (P_N,r ) &=& 
   \sum_{i,j} c_{ij}^J W_{A}^{(\pm)~(i) B}(P_N) O^{(j)}(P_N,r ) ,
\label{EEq24}
\end{eqnarray}
In this work we have concerned only with the lowest S-wave states of 
HL-quark mesons. For these states the expansion (\ref{EEq24}) is not 
necessary and their internal WF are given as a direct product of the spinor 
WF(Eqs.(\ref{eq:W+,U,C}) and (\ref{eq:W-,U,C})) and the S-wave oscillator-function 
$O_{S}(P_{N},r)$(Eq.(\ref{eq12-B})). For the excited states it is effective to use, 
as bases of expansion Eq.~(\ref{EEq24}), the WF with definite $j_{q}$ in 
HQET. The relation between our LS-bases and those in HQET has been given in 
Ref.\cite{OdaNishimuraIshida}. An investigation along this line of Isgur-Wise function
in semi-leptonic decay processes was made in Ref.\cite{OdaIshidaIshida}.

 
The Bargmann-Wigner spinors $W_{\alpha}^{(\pm) \beta}(P)$, 
aside from the flavor-indices, are defined as the positive/negative 
Fourier amplitudes of solutions $W_\alpha{}^\beta (X)$
of the 
local Klein-Gordon equation as follows:
\begin{eqnarray}
(\frac{\partial^2}{\partial X_{\mu}^{2}}-M^2) \ \ W_{\alpha}{}^{\beta} (X)&=&0
 , \ \\
W_\alpha{}^{\beta} (X)=\sum_{P_{\mu}(P_{0}>0)} 
(
e^{iP\cdot X} W_\alpha^{(+)\beta} (P)&+&e^{-iP\cdot X} W_\alpha^{(-)\beta} (P)
).\nonumber
\label{eq:localKG}
\end{eqnarray}
Then the $W_{\alpha}^{(\pm)\beta}(P)$ are generally given by the 
bi-products(, such as representing the physical situation of relevant 
meson system) of the fundamental vectors of $\tilde{U}(4)_{DS}$ space, 
$W_{\alpha}^{(\pm)}(P)$ and $\bar{W}^{(\pm)\beta}(P)$, as specified in 
the following subsection \ref{sec:D}.

The urciton Dirac spinors, $W_{\alpha}^{(+)}(P)$ and $W_{\alpha}^{(-)}(P)$, 
are, respectively, the positive and negative frequency Fourier amplitudes 
of a single-index BW spinors $W_{\alpha}(X)$. 
For clarity we give the explicit form of them:
\begin{widetext}
\begin{eqnarray}
W_{\alpha}^{(+)}(P) &=& \sum_{s,\bar s=\pm} 
         \left( u_{r=+,s,\alpha}(P) +  v_{\bar{r}=-,\bar{s},\alpha}(P) \right)   ,\\ 
W_{\alpha}^{(-)}(P) &=& \sum_{\bar s,s=\pm}
         \left( v_{\bar{r},\bar{s},\alpha}(P) +  u_{r=-,s,\alpha}(P)  \right)  ,\ \ \\
u_{+,s}(P)_{\alpha}&=&\left(
\begin{array}{c}
ch\theta \chi^{(s)}\\
sh\theta {\bf n}\cdot \mbox{\boldmath $\sigma$} \chi^{(s)}\\
\end{array} 
\right)_{\alpha},\ \ \ \
u_{-,s}(P)_{\alpha}=(-\gamma_5)_{\alpha}{}^{\alpha'} u_{+,s}(P)_{\alpha'},
 \ (-\gamma_5 = \rho_1),\\
{\bar v}_{+,\bar{s}}(P)^{\beta}&=&
(sh\theta \chi^{(\bar{s})\dagger}
{\bf n}\cdot \mbox{\boldmath $\sigma$},
-ch\theta \chi^{(\bar{s})\dagger})^{\beta},\ \ \ \
 \bar{v}_{-,\bar{s}}(P)^{\beta}={\bar v}_{+,\bar{s}}(P)^{\beta'}
 (\gamma_5)_{\beta'}{}^{\beta},
\label{eq:explicit-form}
\end{eqnarray}
\end{widetext}
where $ch\theta =\sqrt{\frac{E+M}{2M}}$ and $sh\theta =\sqrt{\frac{E-M}{2M}}$. 
The $r_3(\bar r_3)$ is the eigen-value of $\rho_3(v) \equiv -iv\cdot\gamma $
($\bar\rho_3 (v) \equiv iv\cdot\gamma $) which reduces to the ordinary 
$\rho_3(\bar\rho_3)$ at the rest frame, where 
$v_\mu \equiv P_\mu /M=(0,0,0,i)$.

The $u_{\pm,s}(P)(\bar{v}_{\pm,s}(P))$ have $r_3=\pm 1 (\bar r_3=\pm 1)$
and (,as was 
mentioned in \ref{sec:1B}) the $u_{+,s}(v_{+,\bar{s}})$ are (relativistic)
Pauli-spinors, which have their correspondents in NRQM; while the
$u_{-,s}(v_{-,\bar{s}})$ are chiral spinors, which have newly appeared
in the $\tilde{U}(12)$-classification scheme.
\subsection{\label{sec:D}
Spinor WF of meson system and chiral transformation
}
In describing the spinor WF of light-quark (LL) meson system,
because of chiral symmetry of the confined light-quarks, the members of
 BW spinors
 with all the combinations of $(r_3,\bar r_3)=(\pm ,\pm )$
are expected to be required in nature. 
Thus the total $W_\alpha{}^\beta (v)$-space 
is equivalent to all the 16 components of Dirac $\gamma$-matrices, 
and it is expanded by them.
Thus, the scalar mesons in $\sigma$-nonet, concerning with $(1)_\alpha{}^\beta = \delta_\alpha{}^\beta$, are naturally
included in the ground-state chiralons. 

The states/mesons described with $(r_3,\bar r_3)=(+ ,+)$, 
which have their correspondents in the conventional 
NR-classification scheme, are Pauli states/Paulons.
The other states/mesons described with $(r_3,\bar r_3)=(+,-),(-,+)$ and $(-,-)$, not appearing in NRQM,
are chiral states/chiralons, since these states are obtained through the 
chiral $-\gamma_5$-transformation
of Pauli states/Paulons, where the value of $r_3$ and $\bar{r}_{3}$
 changes as $+ \rightarrow -$. 
They, Paulons and chiralons, form together a linear representaion of chiral symmtery.

In the heavy-light (HL) meson system the states with
$(r_3,\bar r_3) =  (+,+)$ and 
$(+,-)$ are expected to be 
realized, reflecting the physical situation that the HL meson system has 
the non-relativistic
$SU(2)_s$ spin symmetry (the relativistic, chiral symmetry) concerning the constituent 
Heavy quarks (Light quarks).Accordingly the scalar $S$ 
and axial-vector $A_\mu$ 
chiralons with $(r_3,\bar r_3)=(+,-)$, as well as the 
pseudo-scalar $P_s$ and vector $V_\mu$ Paulons 
with $(r_3,\bar r_3)=(+,+)$ are predicted to exist.

The explicit form of $W_{\alpha}^{(+) \beta}(v)$ and 
$W_{\alpha}^{(-) \beta}(v)$ 
for $D(c\bar q)$-mesons is given as follows (where $\tilde{\gamma}_\mu \equiv
 \gamma_{\mu}- v_{\mu}(v \cdot \gamma)$ 
satisfying $P_\mu\tilde\gamma_\mu =0$);
\begin{widetext}
\begin{eqnarray}
W_\alpha^{(+)\beta}(v)&=& \left( U_\alpha^{(+)\beta}(v),~
                    C_\alpha^{(+)\beta}(v) \right) \label{eq:W+,U,C}\\ 
U_\alpha^{(+)\beta} (v) & \equiv & 
\sum_{s,\bar s} u_{+,s,\alpha}^{(c)} (P)
\bar v^\beta_{(\bar q)+,\bar{s}} ({P}) 
=\left( \frac{1-iv\cdot\gamma}{2\sqrt 2} 
\left[ i\gamma_5 D(P) + i\tilde\gamma_\mu D_\mu (P) \right]
\right)_\alpha{}^\beta, \nonumber \\ 
C_\alpha^{(+)\beta} (v) & \equiv & 
\sum_{s,\bar s} u_{+,s,\alpha}^{(c)} (P)
\bar v^\beta_{(\bar q)-,\bar{s}} ({P}) 
=\left( \frac{1-iv\cdot\gamma}{2\sqrt 2} 
\left[ D_0^\chi (P) + i\gamma_5\tilde\gamma_\mu D_\mu^\chi (P) \right]
\right)_\alpha{}^\beta ,
\nonumber
\end{eqnarray}
\end{widetext}
\begin{widetext}
\begin{eqnarray}
W_\alpha^{(-)\beta}(v) &=& \left( U_\alpha^{(-)\beta}(v),~
                  C_\alpha^{(-)\beta}(v) \right)  \label{eq:W-,U,C}\\ 
U_\alpha^{(-)\beta} (v) & \equiv & 
\sum_{s,\bar s} v_{+,\bar{s},\alpha}^{(\bar c)} (P)
\bar u^{\beta}_{(q)+,s} (P) 
=\left( \frac{1+iv\cdot\gamma}{2\sqrt 2} 
\left[ i\gamma_5 \bar D^\dagger (P) + i\tilde\gamma_\mu 
\bar D_\mu^\dagger (P) \right] \right)_\alpha{}^\beta 
, \nonumber \\ 
C_\alpha^{(-)\beta} (v)  &\equiv& 
\sum_{s,\bar s} v_{+,\bar{s},\alpha}^{(\bar c)} (P)
\bar u^{\beta}_{(q)-,s} (P)
=\left( \frac{1+iv\cdot\gamma}{2\sqrt 2} \left[ \bar D_0^{\chi, \dagger} (P)
+ i\gamma_5\tilde\gamma_\mu \bar D_\mu^{ \chi ,\dagger } (P) \right]
\right)_\alpha{}^\beta  .
\nonumber
\end{eqnarray}
\end{widetext}
In Eq.(\ref{eq:W+,U,C}) and Eq.(\ref{eq:W-,U,C}) 
$U_\alpha^{(+)\beta} (v)$ ($C_\alpha^{(+)\beta} (v)$) are 
represented in terms the positive-frequency part of 
pseudoscalar $D$ and
vector $D_\mu$ (scalar $D_0^\chi$ and axial-vector $D_\mu^\chi$),
while 
$U_\alpha^{(-)\beta} (v)$ ($C_\alpha^{(-)\beta} (v)$)
are represented in terms of the negative-frequency part of 
the respective 
$\bar D(q\bar c)$-mesons.
From  Eq.(\ref{eq:W+,U,C}) and Eq.(\ref{eq:W-,U,C}) 
it is evident that Paulons $(D,D_\mu )$ are 
transformed into chiralons $(D_0^\chi ,D_\mu^\chi )$
through the chiral transformation of light anti-quark, 
$W(v) \rightarrow W(v) e^{i\gamma_5 \frac{\lambda^a \alpha^a}{2}}$
($\lambda^a$ being the flavor $U(3)_F$-matrices).
\subsection{\label{sec:E}
Bi-spinor field of \mbox{\boldmath $D(c\bar q)$}-meson 
system
}
The bi-spinor field of $D(c\bar q)$-meson (($\bar{q}= \bar{u}, \bar{d}, \bar{s}$) in the $(c\bar{q})$-system) denoted as $\Phi_D (X,r)$, 
is defined by
\begin{widetext}
\begin{eqnarray}
\Phi_{D,A}{}^{B}(X,r) &=& \int \frac{d^3{\bf P}}{\sqrt{(2\pi )^3 2E}} \left( 
  W_{D,A}^{(+) B}(v)e^{iP\cdot X} + W_{D,A}^{(-) B}(v) e^{-iP\cdot X} \right)\ O(P,r)\label{eq12-B}\\
 &=& \frac{1}{2 \sqrt 2}\left( 1-\frac{\gamma\cdot\partial}{\sqrt\Box}\right)
        \left( i\gamma_5 \phi_{D,a}{}^{b}+i\gamma_\mu \phi_{D_\mu,a}{}^{b}
              +\phi_{D_{0}^{\chi} ,a}{}^{b}
              +i\gamma_5\gamma_\mu \phi_{D_{\mu}^{\chi} ,a}{}^{b}
               \right) _{\alpha}{}^{\beta} O(P,r), 
\nonumber
\end{eqnarray}
\end{widetext}
where the $\phi_R (X)\ (R=D,D_\mu ,D_0^\chi ,D_\mu^\chi )$ represent the local $D$-meson field in the ($c\bar{q}$) system.
The $\bar D(q\bar c)$-meson field is given by its Pauli-conjugate. 
$\Phi_{\bar D}(X,r)=\bar \Phi_D(X,r)=\gamma_4 \Phi_D(X,r)^\dagger \gamma_4$.
The $\Phi_{\bar D}$ is obtained from $\Phi_D$ by replacing $W^{(\pm)}(v)$ with 
$\bar W^{(\pm)}(v)(\equiv \overline{W^{(\mp)}}(v))$, 
and becomes $\Phi_{\bar{D}}=\bar{\Phi}_{D}$ due to Eq.(\ref{eq:cc}).
\subsection{\label{sec:F}
Static \mbox{\boldmath $U(12)$} symmetry embedded in \mbox{\boldmath $\tilde U(12)$} space of representation
}
The equation (\ref{eqYukawa})(, using the abbreviated notation 
$\Box\equiv (\partial/\partial X_{\mu})^{2}$ and $\gamma \cdot \partial
\equiv \gamma_{\mu}\cdot \partial_{X\mu}$,)
 is derived from the following action as
\begin{widetext}
\begin{eqnarray}
{\cal S} &=& \int d^4X d^4r {\cal L}(X,r), \ \ \ \ \ 
{\cal L}(X,r)\equiv
\langle \Phi_{D}(X,r) ( \partial_{X}^2 - {\cal M}^2(r)) 
        \overrightarrow{F}_{U}(X) \bar{\Phi}_{D}(X,r)  \rangle 
,\ \ \ \ \ \ \ \ \label{action}
\end{eqnarray}
\end{widetext}
where the notation $\langle M \rangle$ denotes to take the trace on the 
spinor-flavor indices, and a factor, to be named as unitarizer, is inserted 
between the trace on light-quarks. 
In Eq.(\ref{action}), through the integration,  
this factor $\overrightarrow{F}_{U}(X) \equiv 
\frac{\gamma\cdot\overrightarrow{\partial}}{\sqrt{\Box}}$ and
$\overleftarrow{F}_{U}(X) \equiv 
\frac{\gamma\cdot\overleftarrow{\rm \partial}}{\sqrt{\Box}}$, 
becomes $F_{U}(v)=\mp iv\cdot\gamma$ for $\bar W^{(\pm)}(v)$, leading to the change of signs,
$\bar U^{(\pm )}\rightarrow -\bar U^{(\pm )}$ and $\bar C^{(\pm )}\rightarrow \bar C^{(\pm)}$,
which makes the ${\cal S}$ chiral-invariant.
It reduces, in the meson rest frame, to $\mp\gamma_4$, and the overlapping changes such as
\begin{widetext}
\begin{eqnarray}
\langle  W^{(+)}(v) {\cal O} (-iv\cdot\gamma)\bar W^{(-)}(v)  \rangle
  &\sim& \langle u_c(P) \bar v_{\bar q}(\pm P) {\cal O} (-iv\cdot\gamma)
             v_{\bar q}(\pm P)\bar u_c(P) \rangle  \nonumber\\
  &\rightarrow& \langle u_c({\bf 0})  v_{\bar q}(\pm {\bf 0})^\dagger
        {\cal O}  v_{\bar q}(\pm {\bf 0})\bar u_c({\bf 0})\cdot  \rangle  ,
\label{eqU12}
\end{eqnarray}
\end{widetext}
where ${\cal O}\equiv (\partial_{X}^{2}-{\cal M}^2(r))/\sqrt{\Box}$ 
is the scalar Klein-Gordon operator. The last line of Eq.~(\ref{eqU12})  
is invariant under the static $U(12)$ transformation, 
$v_{\bar q}({\bf 0})\rightarrow e^{i\frac{\lambda^a}{2}\Gamma_i \alpha^a_i}v_{\bar q}({\bf 0})$,
where $\lambda^a$ are flavor $U(3)$ matrices, $\Gamma_i$ are the hermitian Dirac 16 matrices
 and $\alpha^a_i$ are transformation parameters.
Accordingly the action Eq.~(\ref{action})
leads, after integrating on the $d^{4} r$, to the action with 
the bilinear terms of a series of the 
local $D^{(N)}(c\bar q)$-meson systems with degenerate mass $M_N$, 
in conformity with 
the static $U(12)$ symmetry, as (denoting only the ground state mesons)
\begin{widetext}
\begin{eqnarray}
{\cal S}_{{\rm stat}\ U(12)} 
= -\int d^4X  \sum_{R=D,D_0^\chi , D_\nu ,D_\nu^\chi} 
    \left[  \partial_\mu  \phi_R (X) \partial_\mu \phi_{R}^{\dagger}(X) 
             + \phi_R (X) M^2 \phi_{R}^{\dagger}(X)  \right] .
\ \ \ \ \ \ \ \ \ \label{localKG}
\end{eqnarray}
\end{widetext}
Here, it may be worthwhile to note that, in the case of $S_{\tilde{U}(12)}$ 
(without the above factor $F_{U}(X)$ in Eq.(\ref{action}), the Paulon and 
chiralon-terms in Eq.(\ref{localKG}) obtain the oposite signs each oter.
The action Eq.~(\ref{action}) is used for leading to the conserved electromagnetic current\cite{rfCOQM} 
in section {\ref{sec:4}}. 

\section{Mass spectra for ground HL-quark mesons}
\label{sec:Mass spectra} 
In the $\tilde U(12)$-classification scheme the global mass spectra of quark- and anti-quark mesons
in the low-mass region are to be given generally for both Pauli and chiral states, by 
\begin{eqnarray}
M_N^2 &=& M_0^2 + N \Omega ,\  N\equiv 2n+L\   ,   
\label{eq14}
\end{eqnarray}
(where $\Omega$ denotes inverse Regge slope,)
taking into account to reproduce the phenomenologically well-known Regge trajectories for Pauli-states:
In the relevant HL-quark meson systems all ground-state Paulons($P_s,V_\mu$) and chiralons($S,A_\mu$)
are degenerate in the ideal limit, and
they are expected to split with each others
between chiral partners (spin partners) 
by the spontaneous breaking of the chiral symmetry
(the perturbative QCD spin-spin interaction);
From the approximate chiral symmetry$/$HQS regarding the light$/$heavy quark constituents
we can derive the common relations through the $D(c\bar q)$- and $B(b\bar q)$-meson systems
in the $SU(3)$ limit\cite{BH,rf8}.
\begin{eqnarray}
\Delta M^\chi (Q\bar q) & \equiv & M(0^+) - M(0^-) \nonumber\\
&=& M(1^+) -  M(1^-)\ .\ \ (Q=(c,b).)    
\label{eq15}
\end{eqnarray}

The value $\Delta M^\chi (Q\bar s)$ is determined from the experimental mass values of 
Paulons $(D_s(0^-)$ and $D_s^*(1^-)$) and chiralons($D_{s0}^\chi (0^+)$ and $D_{s1}^\chi (1^+)$), as
\begin{widetext}
\begin{eqnarray}
\Delta M^\chi_{J=0} (Q\bar s)
&=& M_{(c\bar s;0^+)}(2317) - M_{(c\bar s;0^-)}(1968) = 349.2{\rm MeV},\nonumber\\
\simeq \Delta M^\chi_{J=1} (Q\bar s)
&=& M_{(c\bar s;1^+)}(2459) - M_{(c\bar s;1^-)}(2112) = 347.2{\rm MeV},\nonumber\\                       
\ \ \   \rightarrow   && \Delta M^\chi (Q\bar s) \simeq 348 {\rm MeV}.
\label{eq6a}
\end{eqnarray}
\end{widetext}
In Ref.\cite{BEH}, the same value is also applied to the splitting in $Q\bar n$ system as 
$\Delta M^\chi (Q\bar n) = \Delta M^\chi (Q\bar s)$.
Here we consider another possibility, taking
\begin{eqnarray}
\Delta M^\chi (Q\bar n) &=& \Delta M^\chi (Q\bar s) \frac{a}{b} \nonumber\\
&=& 348\times \frac{1}{1.44} =242{\rm MeV},
\label{eq6b}
\end{eqnarray}
where $a\propto \langle u\bar u \rangle_{\rm vac} = \langle d\bar d \rangle_{\rm vac} $,
$b\propto \langle s\bar s \rangle_{\rm vac}$ ($\langle u\bar u \rangle_{\rm vac}$ being the quark bilinear scalar
condensates), and we use the values determined by the 
$SU(3)$ linear $\sigma$ model(L$\sigma$M)\cite{CH,Mune}. 
The Eq.~(\ref{eq6b}) is derived from the $SU(3)$ chiral-symmetric Yukawa interaction 
in the next section.

By using the values Eqs.(\ref{eq6a}) and (\ref{eq6b}) and the experimental masses of 
the ground Pauli-states ($P_s, V_\mu$), we can predict the masses of all the 
ground-state ($L=0$) chiralons ($S,A_\mu$) in HL-quark meson systems 
as given in table \ref{tabMass}.
\begin{table*}
\caption{Mass spectra of ground-state($L=0$) $D_s$, $D$, $B_s$ and $B$ meson systems:
For $D_s$ and $D$ systems, the masses with $L=1$ states are also shown.
The masses of Paulons, $D_0^*$ and $D_1$($D_{s0}^*$ and $D_{s1}$), are the predictions by 
Ref.\cite{Godfrey}(Ref.\cite{God}). 
The recently observed $D_{sJ}(2632)$\cite{2632} state is naturally assigned as $P$-wave
$1^-$ chiral state with $j_q^P=\frac{1}{2}^-$ or $\frac{3}{2}^-$. (Tentatively $\frac{3}{2}^-$ case is chosen
in the table.)
The chiral $1^-$ $D_n$ meson mass $M_{D(1^-)}$ is 
predicted by using simple assumption, 
$M_{D(1^-)}=M_{D_1(2422)}+(a/b)(M_{D_{sJ}(2632)}-M_{D_{s1}(2535)})$, 
with $a/b=1/1.44$. 
The masses of the other $P$-wave $c\bar q$ chiral states are 
given by assuming the spin-dependent splittings being the same as those for
$P$-wave $c\bar s$ Pauli states.}
\begin{center}
\begin{ruledtabular}
\begin{tabular}{rcccccccc}
 & \multicolumn{4}{c}{$c\bar s$-meson} & \multicolumn{4}{c}{$c\bar n$-meson} \\
\hline
$L$   & \multicolumn{2}{c}{Paulons} & \multicolumn{2}{c}{chiralons} 
                         & \multicolumn{2}{c}{Paulons} & \multicolumn{2}{c}{chiralons} \\
\cline{2-3}\cline{4-5}\cline{6-7}\cline{8-9}
 0 &  $0^-$ & $D_s(\underline{1968})$  &  $0^+$ & $D_{s0}^\chi (\underline{2317})$
    & $0^-$ & $D(\underline{1870})$    &  $0^+$ & $D_0^\chi (2110)$ \\
   &  $1^-$ & $D_s^*(\underline{2112})$  &  $1^+$ & $D_{s1}^\chi (\underline{2459})$
    & $1^-$ & $D^*(\underline{2010})$    &  $1^+$ & $D_1^\chi (2250)$ \\
 1 &  $0^+$ & $D_{s0}^*(2466)$\cite{God} &  $0^-$ & $D_{s}^\chi (2563)$       
    & $0^+$ & $D_{0}^*(2400)$\cite{Godfrey} &  $0^-$ &  $D^\chi (2420)$  \\   
   &  $1^+$ & $D_{s1}(2536)$\cite{God} &  $1^-$ & $D_{s}^\chi (2633)$      
    & $1^+$ & $D_{1}(2470)$\cite{Godfrey} &  $1^-$ & $D^\chi (2490)$   \\   
   &  $1^+$ & $D_{s1}(\underline{2535})$ &  $1^-$ & $D_{sJ}^\chi (\underline{2632})$\cite{2632}      
    & $1^+$ & $D_{1}(\underline{2422})$ &  $1^-$ &  $D^\chi (2490)$  \\   
   &  $2^+$ & $D_{s2}^*(\underline{2572})$ &  $2^-$ & $D_s^\chi (2669)$      
    & $2^+$ & $D_{2}^*(\underline{2459})$ &  $2^-$ &  $D^\chi (2525)$  \\   
\hline\hline
$L$      & \multicolumn{4}{c}{$b\bar s$-meson}  &  \multicolumn{4}{c}{$b\bar n$-meson} \\
\hline
 0 &  $0^-$ & $B_s(\underline{5369})$  &  $0^+$ & $B_{s0}^\chi (5717)$
    & $0^-$ & $B(\underline{5279})$    &  $0^+$ & $B_0^\chi (5520)$ \\
   &  $1^-$ & $B_s^*(\underline{5415})$  &  $1^+$ & $B_{s1}^\chi (5760)$
    & $1^-$ & $B^*(\underline{5325})$    &  $1^+$ & $B_1^\chi (5565)$ \\
\end{tabular}
\end{ruledtabular}
\end{center}
\label{tabMass}
\end{table*}
As is stated in the introduction, the ground 
$S$-wave chiral $0^+(D_0^\chi)$ and 
$1^+(D_1^\chi)$ mesons, to be discriminated 
from the ordinary $P$-wave $0^+(D_0^*)$ and $1^+(D_1)$
mesons with $j_q^P=\frac{1}{2}^+$, exist in our scheme.
%
However, for the $(c\bar n)$-system, in the present experiments, 
supposing only one $0^+(1^+)$ meson with broad width exists, the two different values of mass and width
for respective mesons are reported:
For $0^+$ $(M,\Gamma)=(2308\pm 36, 276\pm 66)$MeV\cite{Belle,footBelle}, 
$(2405\pm 28,262 \pm 37)$MeV\cite{FocusD,footFocusD}. 
For $1^+$ $(M,\Gamma)=(2427\pm 36, 384\pm 117)$MeV\cite{Belle,footBelle2}, 
$(2461\pm 50,290 \pm 100)$MeV\cite{CleoD,footCleoD2}. 
The mass of $D_0^{*}(D_1)$ seems to be 
inconsistent(somewhat different) in two experiments, 
suggesting that 
%
the $D_0^*(2308\sim 2405)$ ($D_1(2427\sim 2461)$ ) is 
a superposition of two $0^+$ ($1^+$) resonances,
one is chiral $S$-wave $D_0^\chi (2110)$ ($D_1^\chi (2250)$) and the other is  
$P$-wave $D_0^*(2400)$ ($D_1(2470)$)(, of which masses are predictions by NRQM\cite{Godfrey}). 
The reanalyses of experimental data from this viewpoint are required.

Similarly for the $(c\bar s)$ system, 
the chiral $S$-wave mesons $D_{sJ}^{*}(2317)$ and $D_{sJ}(2460)$,
in addition to the $P$-wave Pauli states $D_{s0}^*(2466)$\cite{God} and $D_{s1}(2536)$\cite{God}, 
are predicted to exist
with broad width, a few hundred MeV, since they have the OZI-allowed $DK$ and $D^*K$ open channels.
 
From Table \ref{tabMass}  
we are able to read the following interesting facts on the hyperfine splittings
$\Delta M^{HF}(Q\bar q)$ between the members with the same quark configurations
\begin{eqnarray}
\Delta M^{HF} \equiv M(1^-) - M(0^-) = M(1^+) - M(0^+),
\label{eq16}
\end{eqnarray}
which is derived from Eq.~(\ref{eq15});\\
$\Delta M^{HF}$ is inversely proportional to the heavy-quark masses, and 
independent of the light-quark masses;
\begin{eqnarray}
\Delta M^{HF}(b\bar q) / \Delta M^{HF} (c\bar q) &=& 0.047/0.14 = 0.34 \nonumber\\ 
  \approx  m_c/m_b (=m_{J/\psi}/m_{\Upsilon}) &=& 3.1/9.5=0.33\ .
\label{eq8a}
\end{eqnarray}
$\Delta M^{HF}$ (with the same heavy-quark) is independent of light-quark masses;
\begin{eqnarray}
\Delta M^{HF}(c\bar n) & \approx & \Delta M^{HF} (c\bar s) \approx  140 {\rm MeV} , \nonumber\\
 \Delta M^{HF}(b\bar n) & \approx & \Delta M^{HF} (b\bar s) \approx  50 {\rm MeV} .
\label{eq8b}
\end{eqnarray}
These facts are reasonably understood from the physical situation in the HL-quark meson systems that 
the heavy-quark (light-quark) behaves non-relativistically (relativistically), 
leading to the HQ symmetry(chiral symmetry) concerning the heavy-(light-) constituent quarks.


The recently observed new $c\bar s$-meson $D_{sJ}(2632)$\cite{2632} causes 
another serious problem\cite{Close} in the conventional classification scheme. 
From its decay modes $D_s\eta$ and $D^0K^+$, the most natural quantum number is $J^P=1^-$.
However, its mass seems to be too light to be assigned as radially
excited $2^3S_1$ state in NRQM. In $\tilde U(12)$-classification scheme, $D_{sJ}(2632)$ 
is naturally assigned as a $P$-wave chiral $c\bar s$-meson with $1^-$.
If this is the case, the other $P$-wave chiral mesons are expected to exist in this mass region
$\sim$2600MeV.
The masses of all these $P$-wave chiral $c\bar q$-mesons, predicted by 
using simple assumptions, are given in Table \ref{tabMass}.

\section{Decay properties of $D_s$-mesons}
\label{sec:4}
\subsection{\label{sec:4A}
Pionic decays
}
In order to estimate the absolute magnitude of the width of the observed pionic decays 
$D_{sJ}^{*}(2317) \rightarrow D_s(1968) + \pi^0$ and 
$D_{sJ}(2460) \rightarrow  D_s^{*}(2112) + \pi^0$,
we shall set up the chiral symmetric effective interaction 
Lagrangian\cite{footlocallag}
\begin{eqnarray}
{\cal L}_{ND} &=& -g_{ND} \langle  \Phi_D(X) M(X)  \Phi_{\bar D}(X)  \rangle 
\label{eq25}\\
{\cal L}_{A} &=& \frac{g_A + g_A^\prime}{2} 
\langle \Phi_D(X) (M(X)\stackrel{\leftarrow}{\partial_\mu}\gamma_\mu )
   \overrightarrow{F}_{U}(X) \Phi_{\bar D}(X) \rangle\nonumber\\
 && +\frac{g_A - g_A^\prime }{2} 
 \langle \Phi_D(X) \overleftarrow{F}_{U}(X)
   (\gamma_\mu\partial_\mu  M(X)) \Phi_{\bar D}(X) \rangle ,\nonumber
\end{eqnarray}
where only the Yukawa interaction of the scalar and pseudo-scalar nonets, 
$M \equiv s - i \gamma_5 \phi$ 
$(s\equiv s^a \lambda^a /\sqrt 2 , \phi\equiv \phi^a \lambda^a /\sqrt 2)$ 
as external fields with the light constituent quarks in the HL-meson is taken into 
account.

The interaction Eq.~(\ref{eq25}) consists of the three parts:\\
Firstly the $g_{ND}$ term (Yukawa interaction in non-derivative form) 
gives dominant (compared to the $g_A$ term) contribution 
to the (quark-) spin non-flip processes.
In spontaneous breaking of chiral symmetry,
 $s$ takes the vacuum expectation value $\langle s\rangle_0 = diag\{ a,a,b \}$
which induces the mass-splittings between chiral partners through
the equation $\Delta M^\chi (c\bar n)=2g_{ND}a$ and 
$\Delta M^\chi (c\bar s)=2g_{ND}b$. 
By using SU(3)L$\sigma$M\cite{CH,Mune}, the $a$ and $b$ are related with the pion and kaon
decay constants as $b=(2f_K-f_\pi)/\sqrt 2,a=f_\pi/\sqrt 2$.
From this relation and by using the experimental value of $\Delta M^\chi (c\bar s)=348$MeV,
Eq.~(\ref{eq6a}), the $g_{ND}$ is determined as $g_{ND}=1.84$ giving 
$\Delta M^\chi (c\bar n)=242$MeV, which is used for predicting the masses of chiral 
$D^\chi$ mesons in section \ref{sec:Mass spectra}.

Secondly the $g_A$ interaction  
concerns dominantly (compared to the $g_{ND}$ term) to the spin-flip processes,
$D^{*+}\rightarrow D^0\pi^+$. Thirdly the $g_A^\prime$ interaction concerns the 
spin non-flip processes, $D_0^\chi\rightarrow D\pi$ and $D_1^\chi\rightarrow D^*\pi$.
\begin{table}
\begin{ruledtabular}
\caption{Formula of pionic decay width $\Gamma$ of $D^{*+}$, $D_{0,1}^\chi$ and $D_{s0,1}^\chi$.
The $M(M^\prime )$ is the mass of initial(final) HL-meson, $\omega =-v\cdot v^\prime$
and sin$^2\theta$ is the $\eta$-$\pi^0$ mixing parameter.
The $f_A$, $f_A^\prime$ and $f_A^{\prime s}$ are defined in the text.
}
\begin{center}
\begin{tabular}{lc}
~~~~~~~~~~process&$\Gamma$\\
\hline
\\
$D^{*+}\rightarrow D^0\pi^+$ & $\frac{|{\bf p}|^3}{6\pi MM^\prime}
    \left(\frac{M+M^\prime}{4a}\right)^2 \left( f_A + \frac{2 g_{ND} a}{M+M^\prime} \right)^2$\\
$D_0^\chi \rightarrow D\pi, D_1^\chi\rightarrow D^*\pi$
  & $\frac{3M^\prime |{\bf p}|}{4\pi M}
    \left(\frac{1+\omega }{2}\right)^2 g_{ND}^2 ( 1- f_A^\prime )^2$\\
$D_{s0}^\chi \rightarrow D_s\pi^0, D_{s1}^\chi\rightarrow D_s^*\pi^0$
  & $\frac{M^\prime |{\bf p}|}{2\pi M}
    \left(\frac{1+\omega }{2}\right)^2 g_{ND}^2 ( 1- f_A^{\prime s} )^2 {\rm sin}^2\theta $\\
\end{tabular}
\end{center}
\label{tabF}
\end{ruledtabular}
\end{table}

The formula of the relevant pionic decay widths of $D(c\bar n)$ and $D_s(c\bar s)$ mesons, which are
derived from Eq.~(\ref{eq25}), are collected in Table \ref{tabF}.
The $f_A$, $f_A^\prime$ and $f_A^{\prime s}$ are 
the coefficients of the axial-current of $i\gamma_5\gamma_\mu$-type.
They are related with $g_A$ and $g_A^\prime$ by
$g_A=\frac{f_A}{2a}$ and $g_A^\prime =\frac{f_A^\prime}{2a}=\frac{f_A^{\prime s}}{2b}$,
where the SU(3)-breaking effect  is taken into account through the difference of vacuum expectation values $b/a=1.44$.

The decays of $D_{s0}^\chi\rightarrow D_s\pi^0$ and $D_{s1}^\chi\rightarrow D_s^*\pi^0$ 
are iso-spin violating, and considered to occur by the mixing of intermediate $\eta$ meson with $\pi^0$ meson.
We can estimate phenomenologically the value of mixing parameter sin$^2\theta$, by using the experimental 
branching ratio\cite{pdg} of $D_n(c\bar n)$ meson to the iso-spin violating decay channel as 
\begin{widetext}
\begin{eqnarray}
({\rm sin}\theta )^2_{\rm exp}  \approx
  \frac{ Br ( D_{s}^{*+}  \rightarrow D_s^+ \pi^0 ) (M_{D_s^*}^2 / q^3 ) }{ Br ( D_{s}^{*+}  \rightarrow D_s^+ \gamma ) } / 
  \frac{2 Br ( D^{*+}  \rightarrow D^+ \pi^0 ) (M_{D^*}^2 / q^3 ) }{ Br ( D^{*+} \rightarrow D^+ \gamma ) } 
  = (0.9\pm 0.4)\cdot 10^{-3} ,
\label{eq24}
\end{eqnarray}
\end{widetext}
which seems to be reasonable order of magnitude due to the virtual EM-interaction.

The experimental value\cite{pdg} of 
$\Gamma (D^{*+}\rightarrow D^0\pi^+)=(96\pm 22)$keV$\times 0.677=(65\pm 15)$keV is reproduced by
$f_A=0.521$\cite{footg}, 
which corresponds to $g_A(=f_A/(2a))=3.96$GeV$^{-1}$.

For fixing the value of $g_A^\prime$, we consider two characteristic models:\\
i) $i\gamma_5\gamma_\mu$-model:\ \  $g_A^\prime =-g_A$\cite{footg2}, 
where the strength of $i\gamma_5\gamma_\mu$ coupling
is common to $D^*\rightarrow D\pi$ and $D_0^\chi\rightarrow D\pi$.\\
ii) $\gamma_5\sigma_{\mu\nu}$-model:\ \ This model starts from the
the effective Lagrangian,
\begin{widetext}
\begin{eqnarray}
{\cal L}_{AX}=g_{AX}\langle  W^{(+)}(v)[iq_\mu 
+\sigma_{\mu\nu}(P+P^\prime)_\nu ]
(-iq_\mu)M(q)\bar W^{(-)}(v^\prime )\rangle,
\end{eqnarray}
\end{widetext}
which is motivated by the 
Gordon decomposition of $\bar v(P)i\gamma_5\gamma_\mu v(P^\prime)$.
The ${\cal L}_{AX}$ is equivalent to the ${\cal L}_A$,
 Eq.~(\ref{eq25}), by taking
$g_A=(M+M^\prime)g_{AX}$ and $g_A^\prime =-(M-M^\prime)g_{AX}$.
The experimental $\Gamma (D^{*+}\rightarrow D^0\pi^+)$ leads to the value $g_{AX}=1.02$GeV$^{-2}$,
which gives $f_A=0.521$ and $f_A^\prime =-0.033$ .
The effect of $f_A^\prime$ is negligibly small in this model. 
We predict the values of the relevant pionic decay widths
in two model cases in Table \ref{tabpi}.
\begin{table}
\begin{ruledtabular}
\caption{The predicted values of pionic decay widths of chiral $D_s$ and $D_n$ mesons.}
\begin{center}
\begin{tabular}{lcc}
   & $i\gamma_5\gamma_\mu$-model & $\gamma_5\sigma_{\mu\nu}$-model\\
\hline
$
\Gamma ( D_{s,0}^\chi (2317) \rightarrow  D_s \pi^0 )$& 381$\pm$168keV & 141$\pm$63keV\\
$= \Gamma ( D_{s,1}^\chi (2459)\rightarrow  D_s^* \pi^0 )$ &&\\
$\Gamma ( D_{0}^\chi (2110) \rightarrow  D \pi )$& 313MeV & 144MeV\\
 = $\Gamma ( D_{1}^\chi (2250)\rightarrow  D^* \pi )$ &&\\
\end{tabular}
\end{center}
\label{tabpi}
\end{ruledtabular}
\end{table}
The predicted widths of $D_{sJ}^{*}(2317)/D_{sJ}(2460)$
are consistent with the experimental values\cite{Mikami}
$\Gamma (D_{sJ}^{*}(2317))<4.6$MeV and $\Gamma (D_{sJ}(2460))<5.5$MeV. 

\subsection{\label{sec:rad}
Radiative decay
}

In order to treat systematically 
all the radiative transitions between the HL-mesons
we shall set up the basic EM-interaction Lagrangian as 
\begin{eqnarray}
{\cal S}_I^{EM} &=& \int d^4x_1 d^4x_2 \sum_{i=1,2} j_{i,\mu} (x_1,x_2) A_\mu (x_i) \nonumber\\
&=& \int d^4 X \sum_i J_{i,\mu}(X)A_\mu (X),\nonumber\\
\label{eq31}
\end{eqnarray}
\begin{widetext}
\begin{eqnarray}
 j_{1,\mu} (x_1,x_2) = -i  d  \frac{e_1}{2m_1}
   \langle  \overrightarrow{F}_{U}(x_{1})\bar\Phi \overleftarrow{F}_{U}(x_{1})
   [  \stackrel{\rightarrow}{\partial_{1\mu}} -  \stackrel{\leftarrow}{\partial_{1\mu}} 
                   + g_M i \sigma_{\mu\nu}^{(1)} 
                           (\stackrel{\rightarrow}{\partial_{1\nu}} +  \stackrel{\leftarrow}{\partial_{1\nu}} ) ] 
   \Phi  \rangle  ,\\
 j_{2,\mu} (x_1,x_2) = -i  d  \frac{e_2}{2m_2}
   \langle  \Phi [  \stackrel{\rightarrow}{\partial_{2\mu}} - \stackrel{\leftarrow}{\partial_{2\mu}} 
                   + g_M i \sigma_{\mu\nu}^{(2)}
                           (\stackrel{\rightarrow}{\partial_{2\nu}} +  \stackrel{\leftarrow}{\partial_{2\nu}} ) ] 
   \overrightarrow{F}_{U}(x_{2}) \bar{\Phi} \overleftarrow{F}_{U}(x_{2})  \rangle  ,
\end{eqnarray}
\end{widetext}
where
\begin{eqnarray}
 d&=&2(m_1+m_2).
\end{eqnarray}
Our multi-local current $j_{i,\mu}$ is obtained through the ``minimal substitution"
of $(\partial_{i,\mu} \rightarrow \partial_{i,\mu} - i e_i A_\mu (x_i) )$ to our Lagrangian
${\cal L}$ (see the footnote \cite{foot3}). 
The spin interaction proportional to $g_M$ is introduced following Ref.\cite{Feynman}.

By performing integration on relative space-time coordinates in the first line of 
the above expression Eq.~(\ref{eq31}),
we obtain the effective heavy- and light-quark current of the HL-mesons:
\begin{widetext}
\begin{eqnarray}
J_{1,\mu} &=& \langle  (-iv^\prime\cdot\gamma )W^{(-)}(v^\prime )
\left( e_1(P_\mu +P_\mu^\prime )
 + d \frac{e_1}{2m_1} g_M i \sigma_{\mu\nu} q_\nu \right)
    W^{(+)}(v) \rangle .\nonumber\\ 
J_{2,\mu} &=&\langle  W^{(+)}(v)
\left( (-e_2)(P_\mu +P_\mu^\prime )
 + d \frac{e_2}{2m_2} g_M i \sigma_{\mu\nu} q_\nu \right) 
 (-iv^\prime\cdot\gamma) 
   \bar W^{(-)}(v^\prime ) \rangle .\ \ \ \ \ 
\label{eq311}
\end{eqnarray}
\end{widetext}
This is one of the most simple forms of the covariant generalization of 
convection and spin current in NRQM.
From Eq.~(\ref{eq311}) we can easily check that our effective current $J_{i,\mu}(X)$ is
conserved in the ideal limit, as it should be.

Our effective current has another remarkable feature due to the covariant nature of our scheme.
The spin-current interaction leads to the Hamiltonian
\begin{eqnarray}
{\cal H}_i^{spin} &\equiv&  J_{i,\mu}^{spin} A_\mu \label{eq:intrinsic}\\
&=& \mu_i  i \sigma_{\mu\nu} q_\nu A_\mu 
   = \mu_i  ( -i \rho_1 \mbox{\boldmath$\sigma$}\cdot \mbox{\boldmath$E$}  +   
               \mbox{\boldmath$\sigma$}^{(i)}\cdot \mbox{\boldmath$B$} )
               ,\nonumber
\end{eqnarray}
where $\mu_i \equiv  d \frac{e_i}{2m_i} g_M$.
The ${\cal H}_i^{spin}$ contains the interaction through
the ``intrinsic electric dipole" $-i \mu \rho_1\mbox{\boldmath$\sigma$}$ as well as the one through
the magnetic dipole $\mu \mbox{\boldmath$\sigma$}$.
The ``intrinsic dipole" gives contributions only for the transitions between chiralons and Paulons, 
while does none for the other transitions.

From the effective currents $J_{i\mu}$ in Eqs.~(\ref{eq31}) and (\ref{eq311}), 
we can derive the formula of the relevant radiative decay widths, 
which are given in Table \ref{tab2}.


\begin{table*}
\begin{ruledtabular}
\caption{Formula of radiative decay widths $\Gamma$ for $c\bar q$ mesons:
By using $|{\cal A}|^2$, the $\Gamma$ is given by 
$\Gamma =\frac{\alpha |{\bf q}|^3}{2J_I+1}\frac{(M+M^\prime )^2}{M^2M^{\prime 2}}\ |{\cal A}|^2$,
where $J_I$ is the spin of initial meson and $\alpha =1/137.037$.
Choosing the normal quark moment $(g_M=1)$, $\mu_1=\frac{d}{2}\frac{e_1}{2m_1}$ and
$\mu_2=\frac{d}{2}\frac{e_2}{2m_2}$. $\epsilon =\frac{M-M^\prime}{M+M^\prime}$. 
The $e_1$($-e_2$) is the charge of the first(second) constituent.
For $D_s^+=c\bar s$, $e_1(e_2)=\frac{2}{3}e(-\frac{1}{3}e)$. 
}
\begin{center}
\begin{tabular}{cccccc}
  & $D_{d,s}^{*+} \rightarrow D_{d,s}^+ \gamma$ & $D_{s0}^{\chi} \rightarrow D_{s}^* \gamma$
  & $D_{s1}^\chi \rightarrow D_{s} \gamma$ & $D_{s1}^\chi \rightarrow D_{s}^* \gamma$
  & $D_{s1}^\chi \rightarrow D_{s0}^\chi \gamma$ \\
\hline
$|{\cal A}|^2$ & $(\mu_1+\mu_2)^2$ & $(\mu_1\epsilon +\mu_2)$ & $(\mu_1\epsilon-\mu_2)$
  & $2(\mu_1\epsilon)^2+2\mu_2^2$ & $(\mu_1+\mu_2)^2$\\
\end{tabular}
\end{center}
\label{tab2}
\end{ruledtabular}
\end{table*}
By using Table \ref{tab2} we can predict the widths 
for all the radiative transitions 
between ground state $D_s$ mesons, which are given in Table \ref{tab3}.
We have also shown the comparison with the other models, making reference to Ref.\cite{Colangelo}. 

\begin{table*}
\begin{ruledtabular}
\caption{$\gamma$-decay widths(keV) of the ground state $D_s$ mesons predicted by various models. 
$\pi^0$-decay widths are also shown for reference. 
The constituent quark masses are fixed with the values,
$m_u=m_d \equiv m_n = M_\rho /2$, $m_s=M_\phi /2$, $m_c=M_{J/\psi}/2$.
 }
\begin{center}
\begin{tabular}{cccc@{}c@{}c@{}c@{}cc}
 Processes  & $i\gamma_5\gamma_\mu$ & $\gamma_5\sigma_{\mu\nu}$ 
   & BEH\cite{BEH} & God\cite{GodGam} & CFF\cite{Colangelo} & FR\cite{FR} & CH\cite{CHou} & AG\cite{Azimov}\\
\hline
$D_{s0}^\chi\rightarrow D_s\pi^0$ & 381$\pm$169 & 141$\pm$63 & 21.5 & $\simeq$10 & 7$\pm$1 & 16 
   & 10$\sim$100  & $129\pm 43(109\pm 16)$ \\
$D_{s0}^\chi\rightarrow D_s^*\gamma$ & \multicolumn{2}{c}{19.2} & 1.74 & 1.9 & 0.85$\pm$0.05 & 0.2
   &   & $\le$1.4 \\
\hline
$D_{s1}^\chi\rightarrow D_s^*\pi^0$ & 381$\pm$169 & 141$\pm$63 & 21.5 & $\simeq$10 & 7$\pm$1 & 32 
   &   & $187\pm 73(7.4\pm 2.3)$ \\
$D_{s1}^\chi\rightarrow D_s\gamma$ & \multicolumn{2}{c}{91.6} & 5.08 & 6.2 & $3.3\pm 0.6$ & 
    &  & $\le 5$ \\
$D_{s1}^\chi\rightarrow D_s^*\gamma$ & \multicolumn{2}{c}{57.4} & 4.66 & 5.5 & $1.5$ & 
    &  &  \\
\hline\hline
$D^{*+}\rightarrow D^+ \gamma$ & \multicolumn{2}{c}{1.15} & 1.63 
  & \multicolumn{5}{l}{$\longleftrightarrow$\ \ \ \ \ 1.54$\pm$0.53(experimental value)\cite{pdg}} \\
$D_s^* \rightarrow D_s \gamma$ & \multicolumn{2}{c}{0.33} & 0.43 &  &  &  &  &  \\ 
\hline
$D_{s1}^\chi \rightarrow D_{s0}^\chi\gamma$ & \multicolumn{2}{c}{0.235} & 2.74 &  &  &  &  &  \\ 
\end{tabular}
\end{center}
\label{tab3}
\end{ruledtabular}
\end{table*}

From the results in Table \ref{tab3} we see that our model gives the much 
larger widths for $\gamma$-transition from chiral to Pauli states (1st and 2nd columns), 
compared with the other models. 
Our width is almost the same for transition from Paulon to Paulon (3rd column), 
while it is the much smaller for transition from chiralon to chiralon (4th column),
compared with the prediction by Ref.\cite{BEH}.
This difference comes firstly from 
the above mentioned feature Eq.~(\ref{eq:intrinsic})
of our currents, and secondly from 
the different identification of the relevant mesons:
The narrow $D_s$ mesons are assigned as the conventional $P$-wave excited states 
in the other models,
while they are the $S$-wave chiral states other than the $P$-wave Pauli-states in our scheme.
\subsection{\label{sec:branch}
Branching ratios between radiative and pionic decay widths
}

From the predicted values of pionic and radiative (Table \ref{tab3}) decay widths we obtain
the ratios between them. Making reference again to Ref.\cite{Colangelo}, 
the results are compared with the other models in Table \ref{tab4}.

\begin{table*}
\begin{ruledtabular}
\caption{Ratios of the predicted radiative and pionic decay widths compared with experiments.}
\begin{center}
\begin{tabular}{ccccccc}
Ratio & Experiment & $i\gamma_5\gamma_\mu$ & $\gamma_5\sigma_{\mu\nu}$ & BEH\cite{BEH} & God\cite{GodGam} 
  & CFF\cite{Colangelo} \\
\hline
$\frac{\Gamma (D_{s0}^\chi\rightarrow D_s^*\gamma)}{\Gamma (D_{s0}^\chi\rightarrow D_s\pi^0)}$
  & 0.29$\pm$0.26\cite{Krokovny} & 0.051$^{+0.040}_{-0.016}$ & $0.14_{-0.05}^{+0.10}$ 
  & 0.08 & 0.2 & 0.1 \\
\hline
$\frac{\Gamma (D_{s1}^\chi\rightarrow D_s\gamma)}{\Gamma (D_{s1}^\chi\rightarrow D_s^*\pi^0)}$
  & 0.44$\pm$0.09\cite{pdg} & $0.24_{-0.07}^{+0.19}$ & $0.64_{-0.19}^{+0.52}$
  & 0.24 & 0.6 & 0.5 \\
$\frac{\Gamma (D_{s1}^\chi\rightarrow D_s^*\gamma)}{\Gamma (D_{s1}^\chi\rightarrow D_s^*\pi^0)}$
  & 0.15$\pm$0.11\cite{Krokovny} & $0.15_{-0.05}^{+0.12}$ & $0.40_{-0.12}^{+0.33}$
  & 0.2 & 0.6 & 0.2 \\
\hline
$\frac{\Gamma (D_{s1}^\chi\rightarrow D_s^*\gamma)}{\Gamma (D_{s1}^\chi\rightarrow D_s\gamma)}$
  & 0.40$\pm$0.28\cite{Krokovny} & \multicolumn{2}{c}{0.63} & 0.9 & 0.9 & 0.4 \\
\hline
$\frac{\Gamma (D_{s1}^\chi\rightarrow D_{s0}^\chi \gamma)}{\Gamma (D_{s1}^\chi\rightarrow D_s^*\pi^0)}$
  & $<$0.58\cite{Besson} & $(0.61^{+0.49}_{-0.18})\cdot 10^{-3}$ & $(1.7^{+1.3}_{-0.6})\cdot 10^{-3}$ 
     & 0.13 & &  \\
\end{tabular}
\end{center}
\label{tab4}
\end{ruledtabular}
\end{table*}

As is shown in Table \ref{tab3}, our predicted radiative decay widths of $D_{s0}^\chi$ and $D_{s1}^\chi$ 
is one-order of magnitude larger than the other predictions. 
Concerning pionic decays, 
our $\gamma_5\sigma_{\mu\nu}$-model(, where $g_A^\prime$ is negligibly small,) is essentially 
equivalent to the Ref.\cite{BEH}, where only the $g_{ND}$ and $g_A$ interactions are considered 
in our language.
However, the resulting pionic decay width by Ref.\cite{BEH} is one-order of magnitude smaller 
than our prediction, because the iso-spin violating factor, sin$^2\theta$, used in Ref.\cite{BEH}(,
which is estimated theoretically from the $\eta$-$\pi^0$-mixing angle\cite{GL,Cho,BEH}
as ${\rm sin}^2\theta = \frac{1}{2}\delta_{\eta\pi^0}^2=
1/(2\cdot(2\times 43.7)^2) \simeq 0.65\times 10^{-4}$),
is about one-order of magnitude smaller
than our phenomenological estimation in Eq.~(\ref{eq24}).
As a result, the ratios\cite{BEH} of partial widths of $\gamma$-decay to $\pi^0$-decay 
become similar values to our predictions.  
Only the Ref.\cite{Azimov} except for us predicts the large pionic decay widths, where
another phenomenological estimation of sin$^2\theta (=\frac{2}{3}\epsilon^2)$ is done;
$\frac{2}{3}\epsilon^2 =\frac{2}{3}\frac{{\cal B}(\psi (2S)\rightarrow J/\psi \pi^0)p_\eta^3}
{{\cal B}(\psi (2S)\rightarrow J/\psi \eta)p_{\pi^0}^3}
=\frac{2}{3}((4.07\pm 0.47)\times 10^{-2})^2 =(1.11\pm 0.25)\times 10^{-3})$, which is consistent with
our value.
The measurements of absolute magnitude of the decay widths of $D_{s0}^\chi$ and $D_{s1}^\chi$ 
are required to clarify the situation.

\section{Concluding Remarks}


Through the investigation of this work it may be concluded that
the $D_s(2317)/D_s(2459)$ mesons are shown to be assigned consistently,
in the $\tilde U(12)$-classification scheme, as the scalar and axial-vector chiralons
in the $(c\bar s)$ ground state. If this is the case, the conventional $P$-wave scalar
and axial-vector mesons, $D_{s0}^*$ and $D_{s1}^*$, are expected to exist 
in the higher mass region and  decay with wide widths into $DK$ and $D^*K$, 
respectively. 
The 
$D_{sJ}(2632)$, observed quite recently, is assigned as the $P$-wave 
chiral state with $J^P=1^-$ in the $\tilde U(12)$-classification scheme.

In the $(c\bar n)$ system two set of $0^+$ and $1^+$-mesons are 
predicted to exist in the lower-mass region,
both of which are expected to have wide widths; 
one is $S$-wave chiralons $D_{0,1}^\chi$, and the other is ordinary $P$-wave mesons $D_{0}^*$ 
and $D_1$.
Recent experimental data of $D\pi (D^*\pi)$ mass spectra show a peak-structure with wide width,
which is explained as a single $0^+$($1^+$) meson, $D_0^*(2308\sim 2405)(D_1(2427\sim 2461)$).
However, this peak-structure is considered to come from 
the two interfering resonances of $D_0^\chi$ and $D_0^*$ ($D_1^\chi$ and $D_1$)
in the $\tilde U(12)$-classification scheme, and accordingly
the data are necessary to be reanalyzed along this line.

The radiative decay widths of the relevant $D_{sJ}^{*}(2317)/D_{sJ}(2460)$ mesons 
are predicted to be remarkably larger than those estimated in other works
in our scheme. These are to be checked experimentally.

\begin{acknowledgments}
The authors should like to express their deep gratitude to 
Professor S.~F.~Tuan, whose continual interest in and suggestion on
our new $\tilde U(12)$-classification scheme has given us great encouragements.
They are also grateful to Professor S. Kamefuchi for useful comments 
and warm encouragements.
\end{acknowledgments}

\appendix

\section{\mbox{\boldmath $\tilde U(12)$}-Classification Scheme and Static 
\mbox{\boldmath $U(12)$} Symmmtry}

The argument in this paper is based on a 
covariant classification scheme of hadrons, 
the $\tilde U(12)$-classification scheme, 
which was proposed\cite{rf2} by us several years ago. 
In this scheme composite hadrons have a new type of symmetry extended from
the non-relativistic $SU(6)_{SF}$ spin-flavor symmetry concerned with
light constituent quarks. 
It is defined in the frame of the relevant hadrons all being at rest,
and is called the static $U(12)$ symmetry, $U(12)_{\rm stat}$,
which includes $SU(6)_{SF}$ as a subgroup.
As a matter of fact, such an attempt to generalize the $SU(6)_{SF}$ relativistically
has a long history and the many critical arguments (for example, No-Go theorem\cite{ColemanM})
against it had been appeared.
In this appendix, 
we shall give a somewhat compact review on the $\tilde U(12)$-classification scheme,
and also explain its essential points to overcome the critical points above mentioned.

\subsection{Covariant bi-spinor WF, its charge-conjugation and chiral transformation}

In order to represent the $U(12)_{\rm stat}$ symmetry for the $q\bar q$ meson system,
we must introduce the relativistically covariant wave function(WF),
\begin{eqnarray}
\Phi_A{}^B(x,y),&\ & A=(\alpha ,a),\ B=(\beta ,b),
\label{A1}
\end{eqnarray}
where $\alpha$ and $\beta$ are Dirac indices, $a$ and $b$ are flavor indices, and 
$x$ and $y$ are space-time coordinates of quark and antiquark, respectively.
As is explained in the text, the WF of this form is introduced from rather general requirements
for the composite hadron system that hadrons should have
(i) definite mass and (ii) spin, (iii) definite Lorentz transformstion property, and 
(iv) definite quark-composite structures. We have imaged as a guide the field theoretical
expression for WF as 
\begin{eqnarray}
\Phi_{M,A}{}^B(x,y) &\sim&\langle 0 | \psi_A(x) \bar\psi^B(y) | M\rangle 
\nonumber\\
&&~~~~~~+ \langle M^c | \psi_A(x) \bar\psi^B(y) | 0 \rangle ,
\label{A2}
\end{eqnarray}
where $\psi_A (\bar\psi^B)$ denotes the quark field(its Pauli-conjugate), and
$| M \rangle ( | M^c \rangle )$ denotes the relevant composite meson(its charge-conjugate) state.
The WF of charge-conjugate meson system is represented by
\begin{eqnarray}
\Phi_{M^c,B}{}^A(y,x) &\sim& \langle 0 | \psi_B(y) \bar\psi^A(x) 
| M^c \rangle 
\nonumber\\
&&~~~~~~+ \langle M | \psi_B(y) \bar\psi^A(x) | 0 \rangle .
\label{A3}
\end{eqnarray}
Then, the Pauli-conjugate WF, defined by $\bar\Phi \equiv \gamma_4 \Phi^\dagger \gamma_4$,
of Eq.~(\ref{A2}) satisfy the relation
\begin{eqnarray}
\Phi_{M^c,B}{}^A(y,x) &=& \overline{\Phi_M(x,y)}_B{}^A\ .
\label{A4}
\end{eqnarray}
These relations imply that the total WF $\Phi_A{}^B(x,y)$ of the composite meson system
and its chrage conjugate meson system satisfies the self-conjugate relation,
\begin{eqnarray}
\Phi_{A}{}^B(x,y) &=& \overline{\Phi (y,x)}_A{}^B\ ,
\label{A5}
\end{eqnarray}
where
\begin{eqnarray}
\Phi_A{}^B(x,y) \equiv \sum_M \Phi_{M,A}{}^B(x,y) = \sum_{M^c} \Phi_{M^c,A}{}^B(x,y) .
\end{eqnarray}
The positive frequency part of WF $\Phi_M$ is denoted as 
\begin{eqnarray}
\Psi_{M,A}^{\ (+)B}(x,y) & \sim & \langle 0 | \psi_A(x)\bar\psi^B(y) |M\rangle \ ,
\label{A6}
\end{eqnarray}
which is transformed through charge-conjugation into
\begin{eqnarray}
  & \longrightarrow & \Psi_{M^c,B}^{\ (+)A}(y,x) 
                 \sim \langle 0 | \psi_B(y)\bar\psi^A(x) | M^c \rangle \ ,
\label{A7}
\end{eqnarray}
where $| M^c \rangle = {\cal C}| M \rangle$ and ${\cal C}$ is the charge conjugation operator.
By using the relation
\begin{eqnarray}
{\cal C}^\dagger \psi_B(y) {\cal C}  & = &  
    C_{BB^\prime} {}^t\bar\psi^{B^\prime}(y),\ \ \ C=\gamma_4\gamma_2 \nonumber\\
{\cal C}^\dagger \bar\psi^A(x) {\cal C}  & = &  
   - {}^t\psi_{A^\prime}(x) C^{\dagger A^\prime A},\ \ \ C^\dagger =\gamma_2\gamma_4\ ,
\label{A8}
\end{eqnarray}
we obtain the relation 
\begin{eqnarray}
 \Psi_{M^c,B}^{\ (+)A}(y,x) 
    &=& C_{BB^\prime} \left(^t\Psi_M^{(+)}(x,y)\right)^{B^\prime}{}_{A^\prime} C^{\dagger A^\prime A} ,~~~~~
\label{A9}
\end{eqnarray}
where we use the anti-commutation relation of $\psi_{A^\prime}(x)$ and $\bar\psi^{B^\prime}(y)$.

We derive the similar relation to Eq.~(\ref{A9}) between the negative frequency parts,
$\Psi_M^{(-)}$ and $\Psi_{M^c}^{(-)}$, of WF $\Phi_M$.

Similarly, the chiral $\gamma_5$-transformation is given by using the operator $\chi (\beta )$ as
\begin{eqnarray}
| M \rangle & \longrightarrow & | M^{\chi (\beta )} \rangle = \chi (\beta ) | M \rangle \ .
\label{A10}
\end{eqnarray}
The $\chi (\beta )$ is defined by
\begin{widetext}
\begin{eqnarray}
\chi (\beta )^\dagger \psi_A(x) \chi (\beta ) = U(\beta )_A{}^{A^\prime} \psi_{A^\prime}(x),\ \ 
  U(\beta )=e^{i\frac{\beta^j \lambda^j}{2}\gamma_5},
\label{A11}
\end{eqnarray} 
\end{widetext}
where $\lambda^j$ are flavor $U(3)_F$ matrices and $\beta^j$ are the transformation parameters.
The WF $\Phi_A{}^B$ is transformed as 
\begin{widetext}
\begin{eqnarray}
\Phi_A{}^B(x,y) &=& \sum_M \Phi_{M,A}{}^B(x,y) = 
     \sum_M \left( \langle 0 | \psi_A(x) \bar\psi^B(y) | M \rangle
            +\langle M^c | \psi_A(x) \bar\psi^B(y) | 0 \rangle  \right) \nonumber\\
 & \longrightarrow &  \sum_M \left( \langle 0 | \psi_A(x) \bar\psi^B(y) | M^{\chi (\beta )} \rangle
            +\langle M^{\chi (\beta ),c} | \psi_A(x) \bar\psi^B(y) | 0 \rangle  \right) \nonumber\\
 &=& U(\beta )_A{}^{A^\prime} \Phi_{A^\prime}{}^{B^\prime}(x,y)U(\beta )_{B^\prime}{}^B\ .
\label{A12}
\end{eqnarray}
\end{widetext}

\subsection{Expansion of bi-spinor WF by the complete set}

The Fourier amplitude of $\Phi_A{}^B(x,y)$ is denoted as
$\Phi_A{}^B(p_1,p_2)$ or $\Phi_A{}^B(P;q)$ where $p_1(p_2)$ is the momentum of quark(antiquark),
and $P_\mu (q_\mu )$ is the CM(internal) momentum.
For the light quark $L\bar L$ meson system it
is expanded by the complete set of bi-spinor WF on $\tilde U(4)$ space
$\{\ \Gamma_i\ \}$ (and by the complete set of $q_\mu$-tensors on $O(3,1)_L$ space) as 
\begin{widetext}
\begin{eqnarray}
\Phi_A{}^B(P;q) &\sim& 
  \sum_i \phi_i(P)_{a}^{\ b} \Gamma_{i,\alpha}{}^\beta f_S(P;q) 
 +\sum_i \phi_{i\mu}(P)_{a}^{\ b} \Gamma_{i,\alpha}{}^\beta q_\mu f_P(P;q) \nonumber\\
 && +\sum_i \phi_{i\mu\nu}(P)_{a}^{\ b} \Gamma_{i,\alpha}{}^\beta q_\mu q_\nu f_D(P;q)
 +\cdots  
\label{eqExpand}
\end{eqnarray}
\end{widetext}
where $\phi_i(P)$'s represent the positive and$/$or negative frequency Fourier-amplitudes 
of the respective local meson WF $\phi_i(X)^\prime$s.
The first term without an explicit factor $q_\mu$ describes the $S$-wave states,
while the second(third) term with a factor $q_\mu (q_\mu q_\nu )$ corresponds to
$P$-wave ($D$-wave or radially excited $S$-wave) states.
The summation on the indices $i$ means that on all the 16 component of Dirac $\gamma$ matrices.
The explicit forms of $\Gamma_i$ and $J^{PC}$ quantum numbers of their corresponding $\phi_i(P)$
are
\begin{widetext}
\begin{equation}
\begin{array}{cccccccccc}
\Gamma_i &: &1\ \ \ &i\gamma_5 &i\tilde\gamma_\mu\ \ 
 &\ \ i\gamma_5\tilde\gamma_\mu &-v\cdot\gamma\ \  &\ \ -\gamma_5v\cdot\gamma
 &-i\sigma_{\mu\nu}v_\nu\ \  &\ \ \gamma_5\sigma_{\mu\nu}v_\nu \\
J^{PC} &: &0^{++} &\ 0^{-+} &1^{--}\ \   &1^{++} &0^{+-}\ \  &\ \ \ \ 0^{-+} 
 &1^{--}\ \ &\ \ 1^{+-}\ 
\end{array},
\label{eqGamma}
\end{equation}
\end{widetext}
where $v_\mu =P_\mu /M$ and $i\tilde\gamma_\mu = \tilde\delta_{\mu\nu} i\gamma_\nu$
($\tilde\delta_{\mu\nu} = \delta_{\mu\nu} + v_\mu v_\nu $), 
satisfying $v_{\mu}\tilde{\gamma}_{\mu}=0$.
For the $HL(Q\bar q)$ meson system, the expansion of WF is made in the text.


Here it may be instructive to note that the covariant 
expansion Eq.~(\ref{EEq24}) in the text, leading to Eq.~(\ref{eqExpand}), 
applied to WF of the type Eq.~(\ref{A2}) is rather 
general and is also valid to the other type WF such as 
the BS amplitude
\begin{widetext}
\begin{eqnarray}
\Phi_A{}^B(x_1,x_2)  \sim  \langle 0 | T \psi_A(x_1) \bar\psi^B(x_2) | M 
\rangle
\label{eqBS}
\end{eqnarray}
\end{widetext}
and the gauge-invariant amplitude adopted in Ref.\cite{Suura}, \\
\begin{widetext}
\begin{eqnarray}
\chi_{\alpha\beta}(x_1,x_2) = \langle 0 | \exp[ig \int_{x_1}^{x_2} 
\mbox{\boldmath A}(\mbox{\boldmath x})\cdot d\mbox{\boldmath x}]
q_\alpha (x_1) q^\dagger_\beta (x_2) | M \rangle  .
\label{eqSuura}
\end{eqnarray}
\end{widetext}

\subsection{\mbox{\boldmath $U(12)_{\rm stat}$} 
symmetry and its representation}

The chiral $\gamma_5$-transformation of quark field (\ref{A11})
induces the chiral transformation of meson WF in Eq.~(\ref{A12}), which
changes the members of $\Gamma_i=i\gamma_{5}$ and $1$ to each other.
Thus, the term of pseudoscalar $\pi$ spinor WF with $\Gamma_i=i\gamma_5$ in 
Eq.~(\ref{eqGamma})
is transformed into the one of the scalar $\sigma$ WF with $\Gamma_i=1$ as
\begin{widetext}
\begin{eqnarray}
\phi_{P_s}(P)_a{}^b & (i\gamma_{5})_{\alpha}{}^\beta & f_S(P;q) \longrightarrow
        \phi_S(P)_a{}^b\ (1)_\alpha{}^\beta\ f_S(P;q)\  ,
\label{A17}
\end{eqnarray}
\end{widetext}
both of which are the $S$-wave states, and the former(latter) totally represents
the pseudoscalar(scalar) mesons. There exists another scalar nonet from the $^3P_0$
state, of which WF is given by
\begin{widetext}
\begin{eqnarray}
\Phi_A{}^B(P;q)[^3P_0] &\sim & 
f_{0a}{}^b(P) \tilde\delta_{\mu\nu} (i\tilde\gamma_{\mu})_{\alpha}{}^\beta q_\nu f_P(P;q)\ .
\label{A18}
\end{eqnarray}
\end{widetext}
This WF is contained in the $P$-wave term, proportional to $\phi_{i\mu}(P)$ in Eq.~(\ref{eqExpand}).
Thus, the above two scalar nonets naturally appear in the expansion of our general WF.
The one corresponds to the $^3P_0(f_0(1370))$ nonet which is the Pauli state, appearing also
in NRQM. The other is the $S$-wave state corresponding to $\sigma$ nonet.
The $\sigma$ nonet is degenerate to the $\pi$ nonet in the ideal case with chiral symmetric
phase and they form a linear representation of chiral symmetry.
Here it is notable that actually the observed masses of the members of $\sigma$ nonet are closer
to the $\pi$ nonet and lower than the $f_0(1370)$ nonet. 

Then we notice an interesting possibility 
that a larger group than $SU(6)_{SF}$
including chiral symmetry is realized in hadron spectroscopy.
It is a symmetry $U(12)_{\rm stat} \supset U(4)_{DS}\times U(3)_F$
combining $U(4)_{DS}$ (which is defined in the rest frame of the relevant hadrons) 
for Dirac spinor indices and $U(3)_F$ for light flavors, to be 
called static $U(12)$ symmetry, $U(12)_{\rm stat}$.
Its generators are defined by
\begin{widetext}
\begin{eqnarray}  
\psi_A=\psi_{\alpha ,a} &\rightarrow& 
 \psi^\prime_{\alpha ,a}=\psi_{\alpha ,a}+\delta\psi_{\alpha ,a};\nonumber\\  
\delta \psi_{\alpha ,a} &=& 
  i ( \epsilon^j + \epsilon_5^j\gamma_5 + \epsilon_\mu^j\gamma_\mu 
    + \epsilon_{\mu5}^ji\gamma_5\gamma_\mu + \frac{1}{2}\epsilon_{\mu\nu}^j\sigma_{\mu\nu} 
    )_\alpha{}^\beta (\frac{\lambda}{2}^j)_a{}^b\ \psi_{\beta ,b}\ ,
\label{eqGenerator}
\end{eqnarray}
\end{widetext}
where $\psi_A$ is the quark field, which belongs to 
the fundamental {\bf 12} representation of $U(12)_{\rm stat}$. 
Similarly the antiquark field $\psi^{\dagger B}$ belongs to its conjugate ${\bf 12}^*$ representation.
Here all the {\bf 144} infinitesimal parameters $\epsilon^\prime$s are real,
and  
$U(12)_{\rm stat}$ is a unitary symmetry, where $\psi^\dagger \psi$ is invariant.

As is discussed in the text, 
it is promising 
both phenomenologically and theoretically
that, {\it concerning the light quark{\rm (}antiquark{\rm )} indices $A(B)$ of WF $\Phi_A{}^B$, all the components
of {\bf 12}{\rm (}${\bf 12}^*${\rm )} are treated as physical degrees of freedom.
Accordingly the composite hadrons, including light quarks or antiquarks,
are considered to be classified with the representation 
of $U(12)_{\rm stat}$.}

The light-quark $q\bar q$ mesons in the ground $S$-wave state are classified as
${\bf 12}\times {\bf 12}^*={\bf 144}$ in $U(12)_{\rm stat}$. Their quantum numbers are given
in Eq.~(\ref{eqGamma}), which includes 
the $0^{-+}$ and $1^{--}$ nonets (mixtures of Pauli- and chiral states)
forming ${\bf 6}\times {\bf 6}^*={\bf 36}$ in $SU(6)_{SF}$ 
as well as the $0^{++}$ $\sigma$-nonet (chiral states).
All of them are degenerate in the ideal case of $U(12)_{\rm stat}$
symmetry.\\

The heavy-light $Q\bar q$-meson system is classified following {\bf 6} of $SU(6)_{SF}$ for $Q$
and ${\bf 12}^*$ of $U(12)_{\rm stat}$ for $\bar q$.
The ground state multiplet of $c\bar q$ system includes the $0^+$ and $1^+$ triplets,
which are chiral states appearing newly in $U(12)_{\rm stat}$, 
as well as the $0^-$ and $1^-$ triplets,
which are Pauli-states, also appearing in NRQM.

The light quark $qqq$ baryon system in the $S$-wave states is classified as 
$({\bf 12}\times {\bf 12}\times {\bf 12})_{S}={\bf 364}$, which include baryon and antibaryon.
The {\bf 182} of baryons is decomposed into ${\bf 182}={\bf 56}+{\bf 70}+{\bf 56}^\prime$,
which includes the conventional $({\bf 6}\times {\bf 6}\times {\bf 6})_S={\bf 56}$ in $SU(6)_{SF}$.
Additional ${\bf 70}({\bf 56}^\prime)$ has negative(positive) parity.
The positive parity $N(1440)$, $\Delta (1600)$ and $\Sigma (1660)$ are the candidates of
the ${\bf 56}^\prime$, which are the $S$-wave states and expected to have smaller masses than the 
ordinary $P$-wave baryons. Thus, the existence of chiral ${\bf 56}^\prime$ 
in addition to {\bf 56}
naturally explains the doublings of positive parity states, experimentally confirmed.
The negative parity baryons in {\bf 70} 
decay to the {\bf 56} baryons and the $\pi$-meson octet in $S$-wave.
The overlapping of the WFs is expected to become much larger than those of the decays 
of excited baryons, since both of the initial and final baryons in the relevant decays
are in ground states. Thus, the baryons in {\bf 70} have generally 
very wide widths and are expected to be observed only as backgrounds,
except for the cases of the problematic $\Lambda (1405)$.

Here we should add a following remark.\\
The $U(12)_{\rm stat}$ symmetry transformation mixes the flavor and different spin
components, as the $SU(6)_{SF}$ does. 
In order to define this type of symmetry consistently, 
avoiding an application of No-Go theorem\cite{ColemanM}, 
we must specify the frame of relevant hadron (see also the original approach on this line\cite{Roman}). 
The $U(12)_{\rm stat}$ transformation
is defined in the frame where all hadrons are at rest. This is the meaning of
{\it static} symmetry.
 Accordingly the $U(12)_{\rm stat}$ does {\it not} include
the space-time symmetry for Lorentz boost as a subgroup.
In the original $\tilde U(12)$ theory all of the homogeneous Lorentz group
is included as a subgroup. 
The $\sigma_{\mu\nu}$ defined in Eq.~(\ref{eqGenerator}) 
are interpreted as its generators
only when the $\epsilon_{4i}$ are taken to be pure-imaginary, while they are real in $U(12)_{\rm stat}$.
On the other hand
the chiral $SU(3)_L\times SU(3)_R$ symmetry is included as a subgroup of $U(12)_{\rm stat}$.
Its linear representation is realized in the $U(12)_{\rm stat}$ multiplet.
 
\subsection{\mbox{\boldmath $\tilde U(12)$}-classification scheme and Lorentz covariance}

Our classification scheme of hadrons is based on $U(12)_{\rm stat}$ symmetry,
however, we call it $\tilde U(12)$-classification scheme.
The $\tilde U(12)$ transformation is defined by the same equation as Eq.~(\ref{eqGenerator}),
if we take the transformation parameters $\epsilon_5^j,\epsilon_4^j,\epsilon_{45}^j$ and $\epsilon_{4i}^j$
to be pure imaginary, while the other parameters are real.
The Lorentz boost is included as a subgroup of $\tilde U(12)$ 
since for the boost the $\epsilon_{\mu\nu}^{j\neq 0}=0$ and 
$\epsilon_{4i}^{j=0}(\neq 0)$ is pure imaginary.

The reasons for using the term $\tilde U(12)$ in our scheme are as follows:\\
i) The representation space of $U(12)_{\rm stat}$ is the same as the one of $\tilde U(12)$
at the rest frame of hadrons. Historically the term, ${\bf 144}$(${\bf 364}$)
applied for mesons(baryons), was used in the framework of $\tilde U(12)$ symmetry\cite{Delbourgo}.\\
ii) It is impossible to define $\tilde U(12)$ as exact mathematical group
generalizing $SU(6)_{SF}$ relativisitcally accorging to No-Go theorem\cite{ColemanM}.
Nevertheless, we can extend the $U(12)_{\rm stat}$ representation space,
which is defined at zero velocity of hadrons, to the space with any velocity covariantly
by using the Lorentz booster in $\tilde U(12)$.

This can be done through the following procedures.
The spinor indices for the WF $\Phi_{A=(\alpha ,a)}^{\ B=(\beta ,b)}({\bf v}={\bf 0})$
(which forms the $U(12)_{\rm stat}$ representation space at ${\bf v}={\bf 0}$) are expanded by
the spinors of free-quark type with zero velocity:
\begin{widetext}
\begin{eqnarray}
{\rm for\ index\ \alpha}~~~~~~~~~~~& {\rm for\ index\ \beta} &\nonumber\\
\nonumber\\
u_{+,s}({\bf 0})   = \left(  \begin{array}{c}\chi^{(s)}\\ 0\end{array}\right),
\ \ \ \ \
&  \bar v_{+,\bar s}({\bf 0}) =(0,-\chi^{(\bar s)\prime}), & 
\ \ \  \chi^{(\bar s)\prime}=-i\sigma_2\chi^{(s)*}.\nonumber\\
\nonumber\\
u_{-,s}({\bf 0})=\left( \begin{array}{c}0\\ \chi^{(s)} \end{array}\right),
\ \ \ \ \
&  \bar v_{-,\bar s}({\bf 0}) =(\chi^{(\bar s)\prime},0). &
\label{eqBase0}
\end{eqnarray}
\end{widetext}
Here we should note that $u_-(\bar v_-)$ as well as $u_+(\bar u_+)$ are
required for expansion-bases of Dirac index $\alpha (\beta )$.
These spinors are boosted by free-quark generators $\sigma_{\mu\nu}$ in $\tilde U(12)$ 
into those with non-zero velocity {\bf v} as
\begin{widetext}
\begin{eqnarray}
        u_{+,s}({\bf v})  &=& \left(  \begin{array}{c} ch\theta\chi^{(s)}\\ 
                                   sh\theta{\bf n}\cdot\sigma \chi^{(s)} \end{array} \right),  
          \ \ \   \bar v_{+,\bar s}({\bf v}) 
 =(sh\theta\chi^{(\bar s)\prime}{\bf n}\cdot\sigma ,-ch\theta\chi^{(\bar s)\prime}), \nonumber\\
        u_{-,s}({\bf v})  &=& \left( \begin{array}{c} sh\theta{\bf n}\cdot\sigma\chi^{(s)}\\
                                  ch\theta\chi^{(s)} \end{array}\right), 
          \ \ \   \bar v_{-,\bar s}({\bf v}) 
 =(ch\theta\chi^{(\bar s)\prime},-sh\theta\chi^{(\bar s)\prime}{\bf n}\cdot\sigma ) .\nonumber\\
\nonumber\\
   && ch\theta =\sqrt{\frac{E+m}{2m}}, \ \ \  sh\theta=\sqrt{\frac{E-m}{2m}} .  
\label{eqBasev}
\end{eqnarray}
\end{widetext}
By using these spinors, the $U(12)_{\rm stat}$ representation space 
is extended to any velocity, and the Lorentz covariance is guaranteed for the WF of
composite hadron system. This extended scheme is called $\tilde U(12)$ clasification 
scheme.

For example, $\Phi ({\bf v}={\bf 0})=-i\gamma_5\gamma_4$ corresponds to extra pseudoscalar
state (besides pion state with WF $i\gamma_5$) with $J^{PC}=0^{-+}$, which is expanded by
$u({\bf 0})$ and $\bar v({\bf 0})$ as
\begin{widetext}
\begin{eqnarray}
 \Phi ({\bf v}={\bf 0}) &=& -i\gamma_5\gamma_4=\sum_{s\bar s}\frac{c_{s\bar s}}{i}
   \left( u_{+s}({\bf 0})\bar v_{+\bar s}({\bf 0}) +u_{-s}({\bf 0})\bar v_{-\bar s}({\bf 0})\right)
   \ \ \ \ c_{s\bar s}=\left(\begin{array}{cc} 0 & 1\\ -1 & 0  \end{array}\right) .
\label{eqExtraPs0}
\end{eqnarray}
\end{widetext}
This WF is boosted to $\Phi ({\bf v})$ with the non-zero velocity {\bf v} as
\begin{widetext}
\begin{eqnarray}
 \Phi ({\bf v}) &=& -\gamma_5v\cdot\gamma =\sum_{s\bar s}\frac{c_{s\bar s}}{i}
   \left( u_{+s}({\bf v})\bar v_{+\bar s}({\bf v}) +u_{-s}({\bf v})\bar v_{-\bar s}({\bf v})\right) ,
\label{eqExtraPsv}
\end{eqnarray}
\end{widetext}
where the final form $-\gamma_5v\cdot\gamma$ is also obtained easily from the Lorentz transformation property 
of $\gamma$-matrices. On the other hand
the scalar $\sigma$ WF with $J^{PC}=0^{++}$ is common in any velocity frame as
\begin{widetext}
\begin{eqnarray}
 \Phi ({\bf 0})=\Phi ({\bf v}) &=& 1 =-\sum_{s\bar s} c_{s\bar s}
   \left( u_{+s}({\bf v})\bar v_{-\bar s}({\bf v}) -u_{-s}({\bf v})\bar v_{+\bar s}({\bf v})\right) .
\label{eqS}
\end{eqnarray}
\end{widetext}

\subsection{Urciton spinors and $\rho$-spin}

The spinor WF Eq.~(\ref{eqBasev}) are called urciton spinors. 
The name, urciton(ur-exciton), is used historically in the exciton quark model proposed\cite{urciton} 
by one of the authors 35 years ago, for the purpose of treating multi-quark hadrons
systematically and covariantly. 
In the urciton scheme, each index of the spinor WF is boosted 
by the same velocity as the relevant hadron. 

We should note that in the $\tilde U(12)$-classification scheme
the urciton spinors are purely formal objects to be introduced as expansion-bases
of hadron WF $\Phi_A{}^B$.
The $u_{+s}(\bar v_{+\bar s})$ 
has its correspondents in NRQM
or in free-quark field theory,
while there is no such correspondence for $u_{-s}(\bar v_{-\bar s})$.
As is explained in section \ref{sec:level1}, 
conventionally the spinors $u_-(v_-)$ for quarks(antiquarks) are identified with the
spinors $v_+(u_+)$ for antiquarks(quarks). This is based upon the hole-theory on the 
free quark field theory.
Correspondingly, in NRQM only the NR two-component Pauli-spinors $\chi^{(s)}(\chi^{(\bar s)^\prime})$
for quarks(antiquarks), which becomes equivalent to the upper(lower) two-components
of four component boosted-Pauli spinors $u_+(v_+)$ in the static limit, are applied.  
However, this picture on hole theory and the identification of $u_-=v_+(v_-=u_+)$
is only applicable to the free quarks (or to whole free hadrons), and unable separately to
the indices of confined constituent quarks, coexisting with the other quarks.  

In the original urciton scheme, only $u_+$ and $\bar v_+$ are considered 
as representing physical degrees of freedom. 
However, the equations (\ref{eqExtraPsv})
and (\ref{eqS}) suggest the $u_-$($\bar v_-$) is also realized as the physical degrees of freedom
in composite hadron systems. 
This freedom for light urciton-quark may be dynamically generated, although
it seems a very difficult problem that what is
the explicit field theoretical representation of $u_-$ and $v_-$.

This new $SU(2)$ freedom, describing $u_+$ and $u_-$ ($v_+$ and $v_-$), 
is called $\rho$-spin freedom,
while the ordinary $SU(2)$ spin is called $\sigma$-spin. These names comes from the 
$\rho\times\sigma$ decomposition of Dirac $\gamma$-matrices.
The $u_+(\bar v_+)$ have positive $\rho_3(\bar\rho_3)$, while $u_-(\bar v_-)$ have
negative $\rho_3(\bar\rho_3)$, where $\bar\rho_3=-\rho_3^t$.
The chiral $\gamma_5$-transformation for $\Phi_A{}^B$
is interpreted as the $\rho_1$ transformation for urciton-spinor space,
since $-\gamma_5 u_\pm = u_\mp$ and $\bar v_\pm \gamma_5 =\bar v_\mp$. 
The states including urciton spinors with negative $\rho_{3}(\bar\rho_3)$ components 
are called chiral states,
while the states including the components with only positive $\rho_{3}(\bar\rho_3)$ are called Pauli states.
The Pauli states and chiral states are transformed into each other through chiral transformation, 
and
the linear representation is realized within the members of a 
single $U(12)_{\rm stat}$  multiplet.

\subsection{${\cal S}$ matrix unitarity in \mbox{\boldmath $\tilde U(12)$}
 scheme}

As is explained in the previous subsections,
our $\tilde U(12)$ group is not defined on physical Hilbert space directly, but 
is defined on the urciton space. 
In this connection we should like to note that
our $U(12)_{\rm stat.}$ symmetry seems to be a requirement
on the ``generalized $M$ function''\cite{Weinberg}
proposed for the {\bf 144}-fold way out from the difficulty of $\tilde U(12)$. 
Because of this way our scheme including $SU(6)_{SF}$
becomes consistent with Lorentz covariance. 
In Ref.\cite{Weinberg}, 
the ${\cal S}$ matrix is represented by using the clasical free-quark type spinors as
\begin{eqnarray}
&\langle& {\bf p}_1 n_1 \sigma_1,\cdots| {\cal S} | {\bf p}_2n_2\sigma_2,\cdots \rangle \nonumber\\
 &=& \sum_{N_1,N_2} u_{N_1}^{*}({\bf p}_1n_1\sigma_1)\cdots 
  {\cal M}_{N_1\cdots ,N_2\cdots}({\bf p}_1\cdots ,{\bf p}_2\cdots) \nonumber\\
  &&  \cdot u_{N_2}({\bf p}_2n_2\sigma_2)\cdots \ ,
\end{eqnarray}
where $u_N({\bf p}n\sigma)$ is free Dirac spinor with momentum ${\bf p}$, flavor $n$ and 
spin $j_z=\sigma$. $N=(\alpha ,n^\prime )$ is (spinor,flavor) index.
There 
the $\tilde U(12)$ transformation is defined as acting on indices $N_1\cdots$ and $N_2\cdots$ 
of the ${\cal M}$-function.
The $u_N({\bf p}n\sigma)$ spinors are supposed to be ``not WF in the sense of
representatives of a state in physical Hilbert space. They are purely formal objects,
whose sole purpose in physics is to allow us to define free fields or 
${\cal M}$ functions\cite{Weinberg}. "
The $u_N({\bf p}n\sigma)$ are physically equivalent to our urciton spinors.
Because of this treatment of $\tilde U(12)$ symmetry,
the resulting ${\cal S}$ matrix is shown to be consistent with the Lorentz covariance.

However, in Ref.\cite{Weinberg},
only $u_+$ and $v_+$ (in our terminology) are taken as $u_N({\bf p}n\sigma)$ spinors
for quark and anti-quark, respectively, and
in this prescription
the problem of ${\cal S}$-matrix unitarity in the original $\tilde U(12)$ seems to 
remain still unsolved.
The requirement of unitarity  
leads to the non-linear equation for 
${\cal M}$-matrix as 
\begin{eqnarray}
Im {\cal M} &=& {\cal M} \Sigma {\cal M}^\dagger ,
\end{eqnarray} 
where the $\Sigma$ is given, by using $u_N({\bf p}n\sigma )$, symbolically as
\begin{eqnarray}
\Sigma_{NN^\prime} &=& \sum_{n\sigma} u_N({\bf p}n\sigma ) u_{N^\prime}({\bf p}n\sigma )^\dagger
\nonumber\\
 &=& \sum_s u_{+s, \alpha}({\bf p}) u_{+s,\alpha^\prime}({\bf p})^\dagger\ 
  \delta_{nn^\prime} \nonumber\\
 &=& \left( \frac{1-iv\cdot\gamma}{2}\gamma_4 \right)_{\alpha\alpha^\prime}\ 
  \delta_{nn^\prime}.
\end{eqnarray} 
This $\Sigma$ is apparently non-invariant in $\tilde U(12)$ transformation, and thus, 
$\tilde U(12)$ violates the unitarity. 
In our scheme, the above $\Sigma$ is extended to  
\begin{eqnarray}
\Sigma^\prime_{NN^\prime} &=& \sum_{s} \left( u_{+s}({\bf p})u_{+s}({\bf p})^\dagger
      + u_{-s}({\bf p})u_{-s}({\bf p})^\dagger \right)_{\alpha\alpha^\prime}\ 
 \delta_{nn^\prime} \nonumber\\
      &=& (-iv\cdot\gamma \gamma_4)_{\alpha\alpha^\prime}\ \delta_{nn^\prime}.
\end{eqnarray}
This $\Sigma^\prime$ is also not invariant under $\tilde U(12)$. 
However it reduces, in the static limit $v_\mu \rightarrow ({\bf 0},i)_\mu$, to 
$(\gamma_4 \gamma_4)_{\alpha\alpha^\prime }\ \delta_{nn^\prime } 
=1_{\alpha\alpha^\prime}\delta_{nn^\prime}=\delta_{NN^\prime}$,
and is invariant under $U(12)_{\rm stat}$-transformation. 
Thus, our 
$U(12)_{\rm stat}$
is consistent with the ${\cal S}$ matrix unitarity.



\begin{thebibliography}{99}
\bibitem{RN1} There had been appeared many works along this line since the first 
observation of charmed mesons. As for the recent works, see Ref.\cite{RN2}.
\bibitem{foot1} 
In this connection we should like to refer to the suggestion 
by M. Gell-Mann(mentioned in Ref. \cite{Coleman}) that
 ``a group might be useful for classifying particles,
even it has no connection with the approximate symmetries''.
\bibitem{foot3}
The ${\cal M}^2(r)$ is given, in the framework of COQM\cite{Sakura}, by
${\cal M}^2(r)=d\left(-\frac{1}{2\mu}\partial_r^2+U(r)\right)$ where $\mu$ is the 
reduced mass of the system $\mu =\frac{m_1m_2}{m_1+m_2}$. The $m_1(m_2)$ is the mass
of the 1st(2nd) constituent. The ${\cal L}(X,r)$ is rewritten as
${\cal L}=\Phi_D d\left( \frac{\partial_1^2}{2m_1}+\frac{\partial_2^2}{2m_2}-U(r)  
\right) \Phi_{\bar D}$, which is used in section \ref{sec:4} 
to lead\cite{rfCOQM} to the conserved electromagnetic 
current.
\bibitem{footBelle} The original value is $(M,\Gamma )=
(2308\pm 17\pm 15\pm 28,276\pm 21\pm 18\pm 60)$MeV for $D_0^{*0}$.
\bibitem{footFocusD} Two values are combined: $(M,\Gamma )
=(2407\pm 21\pm 35, 240\pm 55\pm 59)$MeV for $D_0^{*0}$
and $(2403\pm 14\pm 35,283\pm 24\pm 34)$MeV for $D_0^{*+}$.
\bibitem{footlocallag} Here we give only the form of local Lagrangian, 
effective in the low energy region, 
which are reduced from the original multi-local one. 
The interaction with the heavy quarks may be negligibly small due to OZI rule.
\bibitem{footg} We select one of the two choices, where the $g_A$ amplitude 
interferes constructively with $g_{ND}$ amplitude. 
In case of destructive interference, $f_A=-0.646$.
\end{thebibliography}
\end{document}